\documentstyle[12pt,aaspp4]{article}

\received{27 March 2001}
\accepted{01 August 2001}

\begin{document}

\def\sun{$_{\odot}$}
\def\um {$\mu$m}
\def\as {{$^{\prime\prime}$}}
\def\am {{$^\prime$}}
\def\deg{{$^\circ$}}

\title{Infall, Outflow, Rotation, and Turbulent \\ 
Motions of Dense Gas within NGC 1333 IRAS 4}

\author{James Di Francesco\altaffilmark{1}, Philip C. Myers, David J. Wilner,\\
Nagayoshi Ohashi\altaffilmark{2}, and Diego Mardones\altaffilmark{3}}
\affil{Harvard-Smithsonian Center for Astrophysics \\
60 Garden St., MS 42, Cambridge, MA 02138, U.S.A.}
\altaffiltext{1} {currently at the Radio Astronomy Laboratory, 601 Campbell Hall, University of California, Berkeley, Berkeley, CA, 94705-3411 U.S.A.}
\altaffiltext{2} {currently at the Institute of Astronomy and Astrophysics, Academia Sinica, P.O. Box 1-87, Nankang, Taipei 11529, Taiwan, R.O.C.}
\altaffiltext{3} {currently at the Departamento de Astronom\'ia, Universidad de Chile, Casilla 36-D, Santiago, Chile}

\begin{abstract}

Millimeter wavelength observations are presented of NGC 1333 IRAS 4, a 
group of highly-embedded young stellar objects in Perseus, that reveal 
motions of infall, outflow, rotation, and turbulence in the dense gas 
around its two brightest continuum objects, 4A and 4B.   These data have 
finest angular resolution of $\sim$2\as\ (0.0034 pc) and finest velocity 
resolution of 0.13 km s$^{-1}$\/.

Infall motions are seen from inverse P-Cygni profiles observed in 
H$_{2}$CO 3$_{12}$-$2_{11}$ toward both objects, but also in CS (3-2) 
and N$_{2}$H$^{+}$ 1--0 toward 4A, providing the least ambiguous evidence 
for such motions toward low-mass protostellar objects.  Outflow motions 
are probed by bright line wings of H$_{2}$CO 3$_{12}$-$2_{11}$ and CS 
(3-2) observed at positions offset from 4A and 4B, likely tracing dense 
cavity walls.  Rotational motions of dense gas are traced by a systematic 
variation of the N$_{2}$H$^{+}$ line velocities, and such variations are 
found around 4A but not around 4B.  Turbulent motions appear reduced with 
scale, given N$_{2}$H$^{+}$ line widths around both 4A and 4B that are 
narrower by factors of 2 or 3 than those seen from single-dish observations.  
Minimum observed line widths of $\sim$0.2 km s$^{-1}$\/ provide a new low, 
upper bound to the velocity dispersion of the parent core to IRAS 4, and 
demonstrate that turbulence within regions of clustered star formation 
can be reduced significantly.

A third continuum object in the region, 4B\am, shows no detectable
line emission in any of the observed molecular species.

\end{abstract}


\clearpage
\section{Introduction}

Star formation involves both coherent and incoherent gas motions within 
molecular clouds.  Coherent motions are those with an implicitly common 
directionality, and include the infall, outflow, and rotational motions of 
molecular gas associated with extremely young protostellar objects (Mardones 
et al.\/ 1997, henceforth M97); Gregersen et al.\/ 1997; Bontemps et al.\/ 
1996; Ohashi et al.\/ 1997).  Incoherent motions have no directionality and 
appear random, and include thermal motions within molecular gas and nonthermal 
motions of turbulence (i.e., randomly propagating MHD waves; Arons \& Max 
1975).

Understanding gas motions is crucial to understanding the overall flow of 
mass during the star formation process, and this is typically attempted by 
analyzing molecular gas line profiles.  In practice, however, different 
types of motion can be difficult to distinguish because similar line profiles 
can be produced when different molecular gas motions are projected along 
the line-of-sight.  For example, asymmetrically-blue optically-thick lines 
relative to symmetric optically-thin lines have been interpreted recently 
as evidence for infall motions in several molecular cloud cores (e.g., see 
Leung \& Brown 1977; Zhou et al.\/ 1993).  However, outflow or rotational 
motions of molecular gas can also produce such a signature, especially at 
velocities that are low relative to the systemic velocity (see Adelson \& 
Leung 1988; Cabrit \& Bertout 1986).  In addition, thermal and nonthermal 
motions each act to broaden line widths, and together can potentially obscure 
signatures of other motions.

Analysis of the gas motions associated with star formation is further 
complicated by the fact that stars form in groups.  For example, young stars 
are found by number more in close proximity within massive, turbulent cloud 
cores (as in Orion, Perseus, or Ophiuchus) than in isolation in less-massive, 
non-turbulent cores (as in Taurus or Lupus; see Lada, Strom, \& Myers 1993).  
Given the finite resolutions of millimeter telescopes, the apparent proximity 
on the sky of several protostellar objects within a group can confuse the 
relationships between observed molecular gas motions and individual objects, 
especially in more distant regions.

Millimeter interferometers can minimize the problems of ascribing various
types of motion to gas within regions where groups of young stellar objects 
are forming, by providing high-resolution observations of judiciously-chosen
molecular lines.  For example, lines observed at high resolution can yield 
profiles that are less ambiguous in interpretation.  Furthermore, the high 
resolution can allow a better determination of the spatial association of 
gas with specific objects in a crowded field.  In addition, high-resolution 
observations can reveal strikingly the abundance contrasts between outflowing
and non-outflowing gas, so observing species particularly abundant in either 
of these components can better trace its specific motions. 

In this paper, we present millimeter interferometer observations of the
NGC 1333 IRAS 4 region in the L1430 core of Perseus to probe the individual 
coherent and incoherent motions associated with a small group of protostellar 
objects.  NGC 1333 itself is a region forming many protostellar groups, and
has a 350 pc distance\footnote{A distance to NGC 1333 of 220 pc has been 
suggested by \v Cernis 1990, but we assume the 350 pc value for this paper.} 
(Herbig \& Jones 1983).  

IRAS 4 consists of 4 (or more) very-embedded objects located within a 30\as\ 
(0.05 pc) radius, all $\sim$3\am\ (0.3 pc) southeast of SVS 13.  IRAS 4 was 
discovered by Jennings et al.\/ (1987) as a single IRAS source, and later 
resolved by Sandell et al.\/ (1991) into 4A and 4B, separated by 30\as.  
Although 4A and 4B together were considered one of the first ``protobinaries" 
ever detected (see Mathieu 1994), each object is itself multiple.  IRAS 4A was 
resolved into 4A and 4A\am, with 2\as\ separation, by Lay, Carlstrom \& Hills 
(1995), and IRAS 4B was resolved into two objects, 4B and 4B\am, with 10\as\
separation, by Looney, Mundy \& Welch (2000).  The 4B object itself also may 
be a binary or triple system of $<$0\farcs5 separation, although this remains 
uncomfirmed.  The 4B\am\ object has been named differently by other groups 
who detected it contemporaneously, i.e., 4BII by Smith et al.\/ (2000), 4BE 
by Sandell \& Knee (2000), and 4C by Looney et al.\/ and Choi, Panis \& Evans 
(1999).  We name the object 10\as\ east of 4B as 4B\am\ because 4C was used by 
Rodriguez, Anglada \& Curiel (1999) and Sandell \& Knee to name a millimeter 
object $\sim$40\as\ east-by-northeast of 4A.  The spectral energy distributions 
(SEDs) of 4A and 4B are both very red, suggesting that IRAS 4 consists of 
several ``Class 0" objects, each at the earliest observed stage of protostellar 
evolution (i.e., $t$ $\approx$ 10$^{4}$ years or less; see Andr\'e, 
Ward-Thompson \& Barsony 1993).  

IRAS 4 has been identified previously as a site of outflow and possibly 
infall in a turbulent region forming clusters.  Using single-dish observations 
of numerous millimeter and submillimeter emission lines, Blake et al.\/ (1995,
henceforth B95) probed abundance depletions and enhancements in the gas 
associated with 4A or 4B.  Low velocity line emission (i.e., at $\Delta V$ 
$\approx$ 1--2 km s$^{-1}$\/ relative to the systemic velocity) revealed a 
non-outflowing gas component around each source, with abundance depletions 
by factors of 10--20 (relative to dust) due to the adsorption of molecules 
onto grains at the high densities ($n$ = 2 $\times$ 10$^{6}$ cm$^{-3}$) and 
low temperatures ($T_{k}$ = 20-40 K) of the gas surrounding each object.  The 
profiles of some optically-thick lines seen by B95 were asymmetrically-blue, 
suggesting these gas components were infalling.  Outflowing gas around each 
source was also detected from line wings seen at higher relative velocities.  
Qualitative differences in the outflows were found between sources, e.g., 
the 4A outflow extended 4\am\ tip-to-tip in CO and 1\am\ tip-to-tip in CS, 
but the 4B outflow extended only $<$20\as\ in either tracer.  Despite these 
differences, however, abundance enhancements (relative to CO) were found in 
both outflows for several molecules including H$_{2}$CO, CS, and CH$_{3}$CO, 
likely due to the return of grain mantles to the gas phase from grain-grain 
collisions within outflow shear zones.

Further evidence for infall motions at 4A and 4B were obtained from the 
asymmetrically-blue profiles of H$_{2}$CO 2$_{12}$--1$_{11}$ and CS 2--1 
by M97, and HCO$^{+}$ 4--3 and HCO$^{+}$ 3--2 by Gregersen et al.\/ (1997). 
Mardones (1998, henceforth M98) further mapped IRAS 4 in H$_{2}$CO 
2$_{12}$--1$_{11}$ and H$_{2}$CO 3$_{12}$--2$_{11}$ and found 
asymmetrically-blue profiles throughout the region, but the most extreme 
self-absorptions were found at the positions of 4A and 4B (see also Mardones
et al. 2001).  Choi, Panis \& Evans mapped IRAS 4 in HCO$^{+}$ 1--0 and 
HCN 1--0, and again found asymmetrically-blue emission profiles at the
positions of each protostellar object.  In addition, small redshifted 
absorption dips were seen in their HCN spectrum toward 4A, arising from
infalling gas along the line-of-sight.

We describe our millimeter interferometer observations in \S 2.  A description 
of the H$_{2}$CO 3$_{12}$--2$_{11}$, CS 3--2, and N$_{2}$H$^{+}$ 1--0 data 
is presented in \S 3.  Discussion of the infall and outflow motions revealed by 
H$_{2}$CO and the rotational and turbulent motions revealed by N$_{2}$H$^{+}$
follows in \S 4.  In \S 5, a concluding summary is presented.  An Appendix 
describes the continuum data and the relative millimeter SEDs of 4A, 4B, and 
4B\am.

\section{Observations}
IRAS 4 was observed between July 1997 and March 1998 at the IRAM\footnote{The
Institut de RadioAstronomie Millim\'etrique (IRAM) is an international 
institute for research in millimeter astronomy, and is supported by the CNRS 
(Centre National de la Recherche Scientifique, France), the MPG (Max Planck 
Gesellschaft, Germany), and the IGN (Instituto Geografico Nacional, Spain).} 
Plateau de Bure Interferometer (PdBI), NRO\footnote{The Nobeyama Radio 
Observatory (NRO) is a branch of the National Astronomical Observatory, an 
interuniversity research institute operated by the Ministry of Education, 
Science, and Culture, Japan.} Nobeyama Millimeter Array (NMA), and the 
NRAO\footnote{The National Radio Astronomy Observatory (NRAO) is a facility 
of the National Science Foundation, U.S.A., operated under cooperative 
agreement by Associated Universities, Inc.} Very Large Array (VLA).  Table 1 
lists the wavelengths, frequencies, configurations, and dates of observations 
made with each array, as well as the synthesized beam FWHMs and the 1 $\sigma$ 
(rms) continuum sensitivities obtained from combining all respective tracks.

\subsection{Primary Observations} 

\subsubsection{PdBI Data}

At the IRAM PdBI, H$_{2}$CO 3$_{12}$--2$_{11}$ at 225.7 GHz was observed to 
trace infall or outflow motions in the region, following M97 or B95 but using 
other H$_{2}$CO transitions and at higher resolution.  Dense gas can be traced 
by this line, as its critical density $n_{cr}$ $\approx$ 7 $\times$ 10$^{6}$ 
cm$^{-3}$ (from Green 1991), but H$_{2}$CO may be depleted in dense gas 
surrounding protostellar objects and enhanced in outflows (see B95).  In 
addition, N$_{2}$H$^{+}$ 1--0 at 93.17 GHz was observed to trace 
non-outflowing dense gas in IRAS 4, following M97 but at higher resolution.  
This line is abundant in star-forming regions (e.g., Womack, Ziurys \& Wyckoff 
1992) and has also a high critical density, i.e., $n_{cr}$ $\approx$ 2 $\times$ 
10$^{5}$ cm$^{-3}$.  Furthermore, N$_{2}$H$^{+}$ is observed and predicted to 
remain relatively abundant in dense cores with $n$ $\approx$ 10$^{4-5}$ 
cm$^{-3}$ (Benson, Caselli, \& Myers 1998, Bergin \& Langer 1997), but is 
possibly depleted, rather than enhanced, by outflow chemistry (see Bergin, 
Neufeld \& Melnick 1998).  Continuum emission at $\lambda$ = 3.2 mm and 
$\lambda$ = 1.3 mm was observed to locate each protostellar object in IRAS 
4 from associated warm dust. 

The 1 mm and 3 mm bands were observed simultaneously.  Receiver 2 was tuned 
to observe H$_{2}$CO 3$_{12}$--2$_{11}$ over 230 channels in a 20 MHz wide 
band, yielding a channel spacing of 0.10 km s$^{-1}$\/ or a velocity resolution 
of 0.13 km s$^{-1}$\/.  The remaining 2 correlator units were assigned to 
observe continuum emission at $\lambda$ = 1.3 mm over a 300 MHz wide band.
Receiver 1 was tuned to observe N$_{2}$H$^{+}$ 1--0 over 230 channels in a 
10 MHz wide band, allowing all 7 of its hyperfine components to be sampled 
with a channel spacing of 0.13 km s$^{-1}$\/, or a velocity resolution of 
0.16 km s$^{-1}$\/.  The other 2 correlator units were combined to observe 
continuum emission at $\lambda =$ 3.2 mm in a 300 MHz wide band.  

Each track with the IRAM PdBI consisted of 3 pointings centered on 4A, 4B, 
and a position halfway in-between, interleaved with observations of 0415+379 
(3C 111) or 0316+413 (3C 84) for phase and flux calibration.  (The sources 
2230+114 and MWC 349 were also observed as secondary flux calibrators.)  With 
primary beam FWHMs of $\sim$54\as\ and $\sim$22\as\ at 93.17 GHz and 225.7 
GHz respectively, this pointing strategy allowed the construction of an 
over-sampled mosaic at $\lambda =$ 3.2 mm and an under-sampled mosaic at 
$\lambda =$ 1.3 mm.  Each track was calibrated in Grenoble, France using 
the IRAM software package CLIC.  Data from each pointing were combined in 
the visibility plane and inverted simultaneously, and the resulting mosaics 
cleaned using the IRAM software package MAPPING.  Further analysis was 
carried out using the MIRIAD, CLASS, and AIPS software packages.  

\subsection{Secondary Observations} 

\subsubsection{NRO NMA Data}

The CS 3--2 line at 146.97 GHz was observed at the NRO NMA in order to 
trace dense gas, as its $n_{cr}$ $\approx$ 1.3 $\times$ 10$^{6}$ cm$^{-3}$ 
(Evans 1999).  Continuum emission at $\lambda$ = 2.2 mm and $\lambda$ = 2.0 
mm was observed in order to trace warm dust, improving the SED of each 
source.  

The NMA FX correlator was tuned to observe CS 3--2 at 146.97 GHz, over 1024 
channels in a 32 MHz wide band, for a channel spacing and velocity resolution 
of 0.064 km s$^{-1}$\/.  The NMA Ultra-Wide Bandwidth Correlator was tuned over 
a 1024 MHz wide band to observe simultaneously continuum emission at $\lambda$ 
= 2.2 mm and $\lambda$ = 2.0 mm in the lower and upper sidebands respectively.

Each track with the NRO NMA consisted of 2 pointings centered on 4A and 4B, 
interleaved with observations of 0316+413 (3C 84) for phase and flux 
calibration.  With a primary beam at 146.97 GHz of $\sim$51\as, this pointing 
strategy resulted in slightly under-sampled mosaics at $\lambda$ = 2.2 mm 
and $\lambda$ = 2.0 mm.  Data from each track were calibrated in Nobeyama, 
Japan, using the software package UVPROC2.  Further analysis was carried 
out using the MIRIAD software packages.  For better sensitivity of CS 3--2, 
the NMA data were tapered with a 4\as\ $\times$ 4\as\ FWHM Gaussian to obtain 
a final 6\farcs5 $\times$ 4\farcs4 FWHM synthesized beam.

\subsubsection{NRAO VLA Data}

Continuum emission at $\lambda$ = 6.9 mm and $\lambda$ = 1.3 cm was observed 
with the NRAO VLA to trace dust, extending the SED of each source across the 
entire millmeter wavelength range.  Observations at these wavelengths are 
also important because they set effective limits on the amounts of free-free 
emission within the SEDs of young stellar objects at shorter millimeter 
wavelengths, allowing the amounts of thermal emission solely from dust to 
be characterized more accurately (e.g., see Di~Francesco et al.\/ 1997).

Both bands were observed simultaneously by dividing the VLA into two 
sub-arrays, with 13 antennas observing at $\lambda$ = 6.9 mm (43.34 GHz) 
using Q-band receivers, and 14 antennas observing at $\lambda$ = 1.3 cm 
(22.46 GHz) using K-band receivers, each over 100 MHz wide bands.

The NRAO VLA was pointed at 1 position halfway between 4A and 4B to include 
4A, 4B, and 4B\am\ within the $\sim$62\as\ and $\sim$120\as\ primary beam FWHMs 
of the VLA antennas at $\lambda$ = 6.9 mm and $\lambda$ = 1.3 cm respectively.  
Observations of IRAS 4 were interleaved with observations of 0316+413 (3C 84) 
for phase calibration approximately every 10 minutes and X-band reference 
pointing checks approximately every 60 minutes.  Observations of 0134+329 (3C 
48) and 1328+307 (3C 286) were made for flux calibration at both wavelengths.  
Data from this track were calibrated in Socorro, U.S.A., and further analyzed 
using the software package AIPS.

\section{Results} 

Figure 1 shows maps of IRAS 4 in continuum emission at $\lambda$ = 1.3 mm, 
$\lambda$ = 2.0 mm, and $\lambda$ = 3.2 mm, as well as maps of the integrated 
intensities of the H$_{2}$CO 3$_{12}$--2$_{11}$, CS 3--2, and N$_{2}$H$^{+}$ 
1--0 line emission (i.e., the ``zeroth-moment" obtained at a given position 
by summing emission from all channels except those containing values $<$~2 
$\sigma$.) 

\subsection{Continuum}

Figures 1a, 1b, and 1c illustrate the relative positions (in projection) 
of protostellar continuum objects in IRAS 4, namely 4A, 4B, and 4B\am\ at 
3 different wavelengths.  These objects were also detected at $>$4 $\sigma$ 
at $\lambda$ = 2.2 mm and $\lambda$ = 6.9 mm, but only 4A and 4B were detected 
at $\lambda$ = 1.3 cm.  Table 2 lists flux densities measured from our data
(also see Appendix.)

Separated from 4A by only $\sim$1\as, 4A\am\ is not detected distinctly from 
4A in our $\sim$2-4\as\ resolution maps, but the continuum emission around 4A 
is extended towards its position.  For this paper, we limit discussion of 4A 
and 4A\am\ as an unresolved pair rather than consider them individually, i.e., 
4A $=$ (4A + 4A\am).  Similarly, we refer to the 4B object as 4B, though it 
may also be multiple.  Furthermore, the nearby source 4C was not observed as
it lay outside the fields-of-view of our PdBI and NMA mosaics, and our VLA 
images did not have sufficient sensitivity to detect it (based on its flux 
densities at $\lambda$ = 3.6 cm and $\lambda$ = 6.0 cm; see Rodriguez et al.\/)

\subsection{H$_{2}$CO 3$_{12}$--2$_{11}$ and CS 3--2}

Figures 1d and 1e show that H$_{2}$CO and CS are distributed similarly around
4A, 4B, or 4B\am, but remarkable differences are found between objects.  Near 
4A, an extended lobe is located to the south-to-southeast and 1 or 2 compact 
spots are located to the north in both H$_{2}$CO and CS.  The southern lobe 
contains 4 local maxima in the H$_{2}$CO data, but only 2--3 maxima in 
the lower-resolution CS data.  Near 4B, 2 single-peaked, compact lobes are 
seen to the north and south, with similar brightnesses in H$_{2}$CO but very 
different brightnesses in CS.  In addition, three less prominent peaks to the 
east, northwest and southeast of 4B are detected in H$_{2}$CO but not in CS.  
Finally, no H$_{2}$CO or CS emission is detected near 4B\am.  In our H$_{2}$CO 
map, this source is located at the mosaic edge, making it unclear if the lack 
of detected H$_{2}$CO near 4B\am\ is real or due to reduced sensitivity.  In 
our CS map, however, 4B\am\ is positioned well within the fields-of-view, but 
no CS is detected near the object.

Figure 2 shows grids of H$_{2}$CO 3$_{12}$--2$_{11}$ spectra from positions 
centered on 4A and 4B, spaced regularly in 2\as\ increments of R.A. or Dec., 
i.e., $\sim$1 FWHM of the synthesized beam.  The velocity range of each panel 
is centered at +6.96 km s$^{-1}$\/, the average central velocity determined 
from N$_{2}$H$^{+}$ 1--0 data of 4A and 4B (see \S 4.3.2 below), which we 
consider here to be the systemic velocity of the IRAS 4 group.  

Figure 2a reveals that at the 4A position (i.e,. the central grid panel), 
an ``inverse P-Cygni" line profile of blue-shifted emission and red-shifted 
absorption is clearly detected.  The absorption feature has a central velocity 
of 7.6 km s$^{-1}$\/ and a width of $\sim$0.9 km s$^{-1}$\/, and is not 
prominent at any other grid location.  The maximum depth of the absorption 
toward 4A is $\sim$9 K, or 1.3 Jy beam$^{-1}$, approximately equal to the 
peak intensity of $\lambda$ = 1.3 mm continuum emission detected at the 
same position (see Table 2).  With single-dish data, M98 detected 
asymmetrically-blue, single-peaked H$_{2}$CO emission with no absorption at 
the 4A position.  North of 4A, the H$_{2}$CO spectra are double-peaked, with 
a dip at $\sim$7.7 km s$^{-1}$\/, and a weak red wing is seen.  South of 4A, 
the H$_{2}$CO spectra are single-peaked, but a strong blue wing is seen.  
The emission spot due north of 4A in Figure 1d almost entirely consists of 
this red-shifted wing emission.  (The apparent H$_{2}$CO emission spot 
northwest of 4A is incidental map-edge noise.)  The extended lobe seen 
south-to-southeast of 4A in Figure 1d almost entirely consists of 
blue-shifted wing emission.

Figure 2b reveals an inverse P-Cygni line profile at the 4B position (at the 
center of this grid, defined similarly to Fig. 2a).  Here, the absorption has 
a central velocity of 7.9 km s$^{-1}$\/ and a width of $\sim$1.0 km s$^{-1}$\/. 
The maximum depth of the absorption feature toward 4B is $\sim$4 K, or 0.57 
Jy beam$^{-1}$\/, again approximately equal to the peak intensity of continuum 
emission at $\lambda$ = 1.3 mm.  In addition, absorption is not prominent 
at positions offset from the continuum object.  M98 also only detected 
asymmetrically-blue, single-peaked H$_{2}$CO emission with no absorption 
below the continuum at the 4B position.  Furthermore, spectra to the north 
of 4B are double-peaked and a red wing is seen, like that seen north of 4A 
and also with a self-absorption at $\sim$7.7 km s$^{-1}$\/.  South of 4B, the 
line is single-peaked and a blue wing is found, similar to the line south of 
4A.  The bright peaks north or south of 4B seen in Figure 1d predominantly 
consist of either red- or blue-shifted wing emission respectively, and the 
secondary features seen northwest and southeast of 4B consist of blue-shifted 
and red-shifted emission respectively.  (The third emission feature seen in 
Figure 1d near 4B is off the eastern edge of the grid in Figure 2b, and has 
both blue- and red-shifted emission.)

Line profiles similar to those found for H$_{2}$CO 3$_{12}$--2$_{11}$ in 
Figure 2 are also seen for the CS 3--2 line at similar positions, including 
an absorption feature at 7.8-8.0 km s$^{-1}$\/ in the CS spectrum at 4A, and a 
possible analogue at 4B.  These data are not shown since they corroborate the 
H$_{2}$CO data, but are noisier and have lower spatial resolution.

\subsection{N$_{2}$H$^{+}$ 1--0}

Figure 1f shows the integrated intensity (zeroth-moment) of N$_{2}$H$^{+}$ 
1--0 emission around the IRAS 4 objects.  These distributions differ 
significantly from those of H$_{2}$CO and CS (Figures 1d and 1e) around each 
object.  However, the spatial distribution of the integrated intensity 
also differs from object to object.  Near 4A, the N$_{2}$H$^{+}$ emission 
is elongated northeast-southwest, extending somewhat symmetrically 10--15\as\ 
around a bisecting channel at position angle (P.A.) $\approx$ 45\deg\ 
containing 4A and 4A\am.  Near 4B, the N$_{2}$H$^{+}$ emission is more 
compact than that near 4A, although it is slightly elongated north-south.  
In addition, N$_{2}$H$^{+}$ emission is coincident with the 4B object, 
unlike at 4A (i.e., no bisecting channel is seen in the zeroth-moment map.)  
Near 4B\am, N$_{2}$H$^{+}$ is not detected, although its position here is 
well within the field-of-view, unlike in the H$_{2}$CO mosaic.

Figure 3 shows grids of spectra of N$_{2}$H$^{+}$ 101--012, at the same 
positions at and around 4A and 4B shown in Figure 2.  (For clarity, only the 
``isolated" hyperfine component of N$_{2}$H$^{+}$ 1--0 is shown.)  Since 
the 2\as\ spacing of panels in Figure 3 corresponds approximately to half-beam 
spacing at the FWHM of the synthesized beam at 93.2 GHz, neighboring panels 
are not independent, as in Figure 2.  (In addition, the velocity range of these 
grid panels differs from that of Figure 2, but remains centered at the same 
systemic velocity.)

Figure 3a reveals (in the central panel) another inverse P-Cygni line profile 
at the 4A position, with absorption of central velocity and width similar to 
those of H$_{2}$CO at the same position, i.e., 7.9 km s$^{-1}$\/ and $\sim$1 
km s$^{-1}$\/ respectively.  (Absorption seen here in adjacent panels is still 
coincident with the continuum object due to the half-beam, non-independent grid 
spacing.)  Unlike the H$_{2}$CO absorption, the peak depth is only $\sim$1.2 
K, or $\sim$50\% the peak continuum intensity of 4A at $\lambda$ = 3.2 mm.  
M98 did not detect absorption of N$_{2}$H$^{+}$ 1--0 in their single-dish 
data at the 4A position.  At positions northwest and southeast of 4A, the line 
is single-peaked in emission only with small shifts in central velocity to 
the red and blue respectively.  No systematic line width variations or wing 
emission is seen.  The bisecting channel at a P.A. $\approx$ 45\deg\ seen in 
Figure 1f is noticeable here from the relatively dim emission found in panels 
on the northeast-southwest diagonal. 

Figure 3b shows that an inverse P-Cygni line profile is {\it not}\/ seen at 
the position of the 4B object, rather a fairly symmetric line is found peaking 
at 7.1 km s$^{-1}$\/, near the systemic velocity.  (Absorption may be present 
at $\sim$8 km s$^{-1}$\/ in the grid panel 2\as\ west of the central position, 
but this is not suggested in any other adjacent, non-independent grid panel.)  
Unlike near 4A, there does not appear to be substantial variations in central
velocity over the grid, but the line is dimmer and broader south of 4B and 
brighter and narrower north of 4B.  Like near 4A, there is no evidence for 
wing emission.  M98 detected a similarly-shaped single-peaked line profile 
at the 4B position.  

\section{Discussion}

\subsection{Distinct Tracers of Specific Motions}

The spectral and spatial distributions of the lines described in \S 3 trace 
quite distinctly specific types of motion within IRAS 4.  Infall is 
traced from the inverse P-Cygni profiles detected in H$_{2}$CO toward both 
4A and 4B (and also in CS and N$_{2}$H$^{+}$ toward 4A alone.)  Outflows from 
4A and 4B are traced distinctly from infall, at positions offset from the 
protostellar objects, from H$_{2}$CO and CS lines with profiles of only bright 
line wings and no line cores.  Rotation of non-outflowing dense gas around 
4A or 4B is traced distinctly from outflow from spatial variations of the 
central velocity of N$_{2}$H$^{+}$, with profiles of only line cores and no 
line wings.  Furthermore, variations of turbulent motions within the 
non-outflowing dense gas about 4A and 4B are traced by spatial variations 
of the line width of N$_{2}$H$^{+}$ in the same profiles.

The observed lines distinctly trace specific motions partly because of 
the differing gas-grain chemistry of the molecular species.  B95 suggested 
H$_{2}$CO and CS were depleted in the non-outflowing gas associated with 4A 
and 4B, due to significant amounts of these molecules adsorbing onto dust 
grains.  Detailed models by Bergin \& Langer (1997) of relative abundances 
in centrally-condensed cores of increasing density support this scenario, 
showing H$_{2}$CO or CS may deplete by $\geq$2 orders of magnitude at densities 
of $n$ $\approx$ 10$^{5}$ cm$^{-3}$ onto grains with H$_{2}$O or CO mantles.  
These models, in turn, also suggest that N$_{2}$H$^{+}$ remains at a constant 
abundance to densities of $\sim$10$^{6}$ cm$^{-3}$, due to the low binding 
energy of N$_{2}$, its precursor molecule, to grain mantles.  Therefore, we 
expect that N$_{2}$H$^{+}$ traces non-outflowing dense gas better than 
H$_{2}$CO or CS in a suitably dense protostellar core.  In contrast, B95 
suggested the enhanced abundances of H$_{2}$CO and CS found in the line wings 
were due to significant amounts of these molecules returning to the gas phase 
from the disruption of grain mantles in the outflows.  Moreover, the 
single-dish data of M97 show the N$_{2}$H$^{+}$ 1--0 components at 4A and 
4B have Gaussian profiles, with no emission at the wing velocities of H$_{2}$CO 
or CS.  Models of molecular abundance variations from post-shock chemistry by 
Bergin et al.  support this contrast, since the increased abundance in outflows 
of H$_{2}$O, returned to the gas phase from grain mantles, appears to rapidly 
deplete N$_{2}$H$^{+}$.  Therefore, we expect H$_{2}$CO or CS traces better 
outflowing gas than N$_{2}$H$^{+}$.  Such molecular abundance contrasts are 
not unique to IRAS 4 and have been observed similarly elsewhere, e.g., the 
single-dish observations of the L1157 core and outflow by Bachiller \& 
P\'erez-Guti\'errez (1997).  

Another important reason our data distinctly trace specific motions is their 
high spatial resolution.  For example, interferometers allow detection of the 
inverse P-Cygni profiles because the brightnesses of the continuum objects in 
the relatively small beams can exceed the brightnesses of line emission from 
intervening, infalling gas along the line-of-sight, leading to red-shifted 
absorption.  (Blue-shifted emission in the inverse P-Cygni profiles arises 
from concentrated gas on the far sides of the protostellar objects.)  In
addition, the spatial filtering properties of the interferometer can further 
distinguish specific motions better than single-dish observations, especially
at low relative velocities, by selectively sampling the kinematics of compact
structures, and resolving out confusing, extended emission.  For example,
outflowing gas is detected more distinctly in interferometer observations of
H$_{2}$CO or CS, beyond abundance differences, because zones of outflowing
dense gas are expected to be more compact, and hence are more easily seen
by an interferometer, than the outer, less-dense layers of a circumstellar
envelope that may still contain abundant H$_{2}$CO or CS.  On the other hand,
interferometers can still detect non-outflowing gas in such an envelope using 
N$_{2}$H$^{+}$ because, unlike H$_{2}$CO or CS, it remains relatively abundant 
in the denser and more-compact inner regions that are less resolved out than
outer regions.  (As described below, N$_{2}$H$^{+}$ may be somewhat depleted 
relative to ``canonical" values in IRAS 4.)

\subsection{H$_{2}$CO Observations}

\subsubsection{Infall}

Detections of red-shifted absorption indicative of infall motions have 
been made primarily toward high-mass young stellar objects, where the 
absorption profiles were seen against the continuum emission of bright HII 
regions in interferometer beams.  For example, red-shifted absorption was 
seen toward G10.6--0.4 in NH$_{3}$ (1,1) and (3,3) by Keto, Ho \& Haschick 
(1987, 1988; with VLA beams of 3\farcs0 FWHM and 0\farcs3 FWHM respectively).  
Similar profiles were also detected toward W51:e2 in NH$_{3}$ (1,1) and (2,2) 
by Ho \& Young (1996; with VLA beams of 2\farcs6--3\farcs8 FWHM), and in 
these transitions plus NH$_{3}$ (3,3) by Zhang \& Ho (1997; with VLA beams 
of 0\farcs2--1\farcs0 FWHM), and in CS 3--2 by Zhang, Ho \& Ohashi (1998; 
with an NMA 1\farcs1 FWHM beam).  Furthermore, an inverse P-Cygni profile 
was noted toward W49A North:G in HCO$^{+}$ 1--0 by Welch et al.\/ (1987; 
with a BIMA $\sim$7\as\ FWHM beam).

Toward low-mass young stellar objects, detections of red-shifted absorption 
against continuum emission have been very rare, the only case being the weak
inverse P-Cygni profile detected in the HCN 1--0 line toward NGC 1333 IRAS 4A 
by Choi, Panis \& Evans with a BIMA $\sim$8--12\as\ FWHM beam.  (Red-shifted
absorption is also suggested in their Figure 1f in the HCN line toward IRAS 
4B, and in the HCO$^{+}$ 1--0 line toward both objects, but they describe only 
the HCN line toward 4A as an inverse P-Cygni profile.)\footnote{An inverse 
P-Cygni profile of the $^{13}$CO (2--1) line toward 4A has been also recently 
detected by Webster \& Welch (2000) with a BIMA 6\farcs7 FWHM beam.}  

Figure 4 shows the H$_{2}$CO 3$_{12}$--$2_{11}$ line at positions of maximum 
continuum intensity from 4A and 4B (enlarged from the central panels of Figures 
2a and 3a.)  These data provide the least ambiguous evidence for infall motions 
toward low-mass young stellar objects.  First, the detection of the absorption 
here is very strong, given the low channel-to-channel noise of these data.  
Second, the excitation requirements of the lower energy level of the H$_{2}$CO 
3$_{12}$--2$_{11}$ transition (i.e., $E_{2_{11}}/k$ = 7.5 K) are higher than 
those of HCN 1--0, suggesting the absorptions are more likely occurring in 
warmer gas associated with the young objects, rather than in a layer of cooler 
gas at slightly higher velocity situated along the line-of-sight (and possibly 
not associated with the young objects.)  In addition, the scales probed by 
these data are smaller than other investigations.  At a distance of 350 pc, 
the linear beam size (FWHM) of the H$_{2}$CO data at IRAS 4 is $\sim$0.003 pc, 
a factor of $\sim$4-6 less than the scales probed by Choi et al., and a factor 
of $\sim$2 less than those probed with inverse P-Cygni profiles at the 
more-distant, higher-mass objects noted above.  

To probe conditions in the absorbing gas, the inverse P-Cygni profiles toward 
4A and 4B were modeled with an enhanced version of the ``two-layer" code used 
by Myers et al.\/ (1996) to reproduce successfully the appearance of spectral 
lines from infalling gas toward several protostellar cores (see also Williams
\& Myers 2000, and Lee, Myers \& Tafalla 2001)  In this version, two layers, 
located along the line-of-sight, each have peak optical depth $\tau_{o}$, 
velocity dispersion $\sigma$, and approach speed $V_{in}$ toward a continuum 
source located in-between.  The continuum source is assumed to be optically 
thick with a Planck temperature $J_{c}$, where $J_{c}$ is related to the 
blackbody temperature $T_{c}$ at frequency $\nu$ by $J_{c}$ = 
$T_{o}$/(exp($T_{o}$/$T_{c}$)-1), with $T_{o}$ = $h\nu/k$, $h$ = Planck's 
constant, and $k$ = Boltzmann's constant.  The continuum source also fills a 
fraction $\Phi$ of the telescope beam area.  The ``front" layer between source 
and observer has a Planck excitation temperature $J_{f}$ while the ``rear" 
layer behind the source has a Planck excitation temperature $J_{r}$.  The rear 
layer is illuminated by the cosmic background radiation of Planck temperature 
$J_{b}$.

For this system, the observed line brightness temperature at velocity $V$ can 
be written as

$$ \Delta T_{B} = (J_{f} - J_{cr})(1-{\rm exp}(-\tau_{f}))+(1-\Phi)(J_{r}-J_{b})(1-{\rm exp}(-\tau_{r}-\tau_{f})), \eqno{(1)} $$

\noindent
where 

$$ J_{cr} = \Phi{J_{c}} + (1-\Phi){J_{r}}, \eqno{(2)} $$ 

\noindent
and,

$$ \tau_{f} = \tau_{o}{\rm exp}{(-(V-V_{in}-V_{LSR})^{2}/2{\sigma}^{2})}, \eqno{(3a)}$$ 

$$ \tau_{r} = \tau_{o}{\rm exp}{(-(V+V_{in}-V_{LSR})^{2}/2{\sigma}^{2})}. \eqno{(3b)}$$

In this model, the line profile can be understood very simply from Equation 1
as the linear sum of a broad positive profile centered on $V_{LSR}$ (i.e., the
right term of RHS) and a narrow negative profile centered on $V_{LSR}$+$V_{in}$ 
(i.e., the left term of RHS).  If the systematic motion of the layers is inward 
(i.e., $V_{in} > 0$\/), Equation 1 implies that the resulting profile shows red 
absorption or ``infall asymmetry."  If the front layer is cooler than the 
continuum source and rear layer by a relatively small temperature differences 
$J_{f}$--$J_{cr}$, the asymmetric profile shows ``self-absorption" but remains 
positive at all velocities.  As this temperature difference increases, the 
profile becomes more asymmetric and its ``dip" goes below the spectral 
baseline. 

Equations 1-3 are essentially the same as Equations 1 and 2 of Myers et al. 
(1996), except that the emission from an optically thick central source is now 
included.  This model is intended to give the simplest possible description 
of a contracting or expanding system which can reproduce the observed line 
profiles.  Therefore, it gives only the crudest picture of the optical depth
and excitation temperature of the absorbing and emitting gas.  However, the
model is still useful for comparing the characteristic velocity $V_{in}$ from 
one line profile to the next, since the ratio $V_{in}$/$\sigma$ is most 
important to setting the line profile for any model of a contracting system 
(Leung \& Brown 1977).

In addition to the observed spectra, Figure 4 shows model inverse P-Cygni 
profiles that successfully reproduce those observed.  The respective model 
parameters are listed in Table 3.  Infall speeds of $V_{in}$ = 0.68 km 
s$^{-1}$\/ and 0.47 km s$^{-1}$\/ are obtained toward 4A and 4B respectively.  
(The blue wing in the 4A spectrum is likely from outflow and so it is not 
fit by the model.)  Using essentially the same two-layer model (but without
a continuum object), M98 found a $V_{in}$ of only 0.11 km s$^{-1}$\/ for the 
IRAS 4 region from the same H$_{2}$CO line.  The difference between these 
results is likely because the spectrum modeled by M98 was obtained with a 
resolution coarser by a factor of $\sim$6 than those shown here, and consisted 
of data averaged over the FHWM of their single-dish integrated intensity map, 
i.e., over a projected distance from the source of $\sim$0.04 pc.  The 
resulting emission spectrum included data from positions offset from 4A and 
4B with gas having possibly lower apparent infall velocities due to projection, 
whereas the inverse P-Cygni profile traces material along the line-of-sight 
to 4A or 4B with minimal projection effects.

The inverse P-Cygni profiles alone cannot demonstrate whether the infalling 
material consists of dense gas individually surrounding 4A or 4B, or from a 
common layer surrounding both objects, since infalling material at projected 
locations between 4A and 4B is not traced.  However, the difference between 
$V_{in}$ toward 4A and 4B is significant in that the range of $V_{in}$ able 
to reproduce well each spectrum is only a few hundredths of km s$^{-1}$\/, 
if all other parameters are kept fixed.  Different values of $V_{in}$ toward 
4A and 4B suggest the absorptions arise in dense gas residing around each 
source rather than in a layer common to both objects, where presumably more 
similar velocities toward 4A and 4B would be found.  

From simple gravitational arguments, the line-of-sight speeds implied by 
the absorptions allow crude estimates of the masses of the 4A and 4B objects, 
assuming the infall speeds $V_{in}$ arise from the velocity gain of gas 
free-falling from rest at $r$ = $\infty$ to $r$ = $r_{in}$, or

$$ M = (V_{in}^{2} r_{in})/2G = 1.2\Biggl({V_{in}\over{1.0~\rm{km~s^{-1}}}}\Biggr)^{2}\Biggl({r_{in}\over{0.01~\rm{pc}}}\Biggr) M_{\odot}. \eqno{(4)} $$

\noindent
To obtain (an upper limit of) $r_{in}$, the mass obtained from Equation 4 
can be equated with that within a sphere of radius $r_{in}$ of mean density 
$n$, i.e., $M = (4/3) \pi r_{in}^{3} n \mu m_{H}$, with $\mu$ equal to the
mean molecular weight (2.32), and $m_{H}$ equal to the mass of hydrogen.  We
assume $n$ = 2.3 $\times$ 10$^{6}$ cm$^{-3}$, the critical density of the 
H$_{2}$CO 2$_{11}$--1$_{10}$ line, a sufficient density for substantial 
population of the 2$_{11}$ energy level of H$_{2}$CO, determined at T$_{k}$ = 
30 K (Green 1991) since B95 estimated a kinetic temperature of non-outflowing 
gas around 4A of 20--40 K.  Accordingly, the values of $V_{in}$ obtained for 
4A and 4B yield values of $r_{in}$ of 0.0099 pc and 0.0068 pc respectively, 
or 5.8\as\ and 4.0\as\ respectively at the assumed distance of IRAS 4.  These 
radii are smaller by factors of 1.4-2.2 than those obtained using integrated
N$_{2}$H$^{+}$ 1--0 emission (see Figure 1f) to trace the projected extents 
of dense gas around 4A and 4B, which is consistent given the lower critical 
density of that transition.\footnote{Non-outflowing H$_{2}$CO in emission, 
i.e., over the velocities traced by N$_{2}$H$^{+}$, is difficult to see around 
4A and 4B, possibly due to low abundances or the relative brightness of the 
outflows.} Using Equation 4, the masses interior to the respective values of 
$r_{in}$ are 0.53 M\sun\ and 0.17 M\sun\ for 4A and 4B, reasonable values 
given the moderate luminosities of these objects and their location within 
a region of low-mass star formation.

Mass accretion rates, $\dot M_{in}$, can be estimated toward 4A and 4B using 
the simple expression: $\dot M_{in} = 4\pi r_{in}^{2} \mu m_{H}nV_{in}$, or

$$\dot M_{in} = 1.7 \times 10^{-4} \Biggl({r_{in}\over{0.01~\rm{pc}}}\Biggr)^{2} \Biggl({n\over{2.3~\times~10^{6}~\rm{cm}^{-3} }}\Biggr) \Biggl({V_{in}\over{1.0 \rm{~km~s^{-1}}}}\Biggr) \rm{~M_{\odot}~yr^{-1}}, \eqno{(5)} $$

\noindent
assuming spherical symmetry of the infall.  With the values of $V_{in}$, 
$n$, and $r_{in}$ described above,  we estimate $\dot M_{in}$ = 1.1 $\times$ 
10$^{-4}$ M\sun\ yr$^{-1}$\/ for 4A, and 3.7 $\times$ 10$^{-5}$ M\sun\ 
yr$^{-1}$\/ for 4B, using Equation 5.  Assuming these rates are constant, 
accretion timescales $t_{acc}$ can be then estimated by dividing the masses 
found above with the accretion rates.  Such a calculation yields $t_{acc}$ 
= 4.7 $\times$ 10$^{3}$ yr and 4.6 $\times$ 10$^{3}$ yr for 4A and 4B 
respectively, remarkably similar and well within the $t_{acc}$ $<$ 10$^{4}$ 
surmised for Class 0 objects. 

If the isothermal sound speed of this gas can be approximated as $V_{rms}$ 
= ($kT_{k}/\mu m_{H}$)$^{1\over{2}}$ and $T_{k}$ $\approx$ 30 K, the values 
obtained here for $V_{in}$ suggest the infall motions toward 4A or 4B are 
supersonic, by factors of 2.1 and 1.4 respectively.  Given that the magnetic 
field strength is poorly known in IRAS 4, however, it is not clear if the 
infall motions are also sub- or super-Alfv\'enic.  From a virial theorem 
argument, Akeson et al.\/ (1996) estimated large magnetic field strengths 
associated with 4A of 5--8 mG, suggesting the infall flows are very 
sub-Alfv\'enic.  However, the summary of actual magnetic field measurements 
(of other various molecular clouds) by Crutcher (1999) suggests these values 
may be too large by factors of $\sim$10-100.  If the infall motions are 
occurring exactly at the Alfv\'en speed (i.e., $V_{A}$ = $B$/($4\pi \mu n 
m_{H}$)$^{1/2}$ with $n$ = 2.3 $\times$ 10$^{6}$ cm$^{-3}$, the required 
magnetic field strengths near 4A and 4B would be 720 $\mu$G and 500 $\mu$G 
respectively.

The two-layer models used to compute the line profiles seen in Figure 4 
neglect the radial variations in density, temperature, and velocity expected 
within protostellar envelopes.  Therefore, these results are too simple to be 
definitive, but provide motivation for more observations and models of inward 
motions.  It will be useful to make more-detailed, Monte-Carlo models of 
radiative transfer through envelopes, of the H$_{2}$CO line from these and 
other Class 0 objects.

\subsubsection{Outflow}

\bigskip
\centerline{\it 4.2.2.1 Outflow Morphologies and Origins}
\medskip

Figure 5 shows channel maps of H$_{2}$CO 3$_{12}$--2$_{11}$ emission near 4A, 
revealing the underlying structure of the blue 4A outflow lobe better than the 
integrated intensity map (Figure 1d).  In channel maps, the blue lobe consists 
of two prominent features south-by-southwest and southeast of 4A.  Increasing 
in relative velocity, emission from the first feature is first seen at 6.5 km 
s$^{-1}$\/, and it peaks at increasing angular separations from 4A in a Hubble 
Law manner reminiscent of other protostellar outflows (see Bachiller 1996).  
At 2.0 km s$^{-1}$\/, however, the angular separations of the peaks of this 
feature shift to the southeast with increasing velocity out to the band edge, 
-10 km s$^{-1}$\/.  The second feature is first seen at 6.8 km s$^{-1}$\/ and 
it moves south and west with increasing relative velocity in a manner less 
consistent with a Hubble Law.  Together, these features resemble cavity walls 
of an outflow seen at increasing relative velocities.  In contrast to Figure 
1d, the channel maps suggest the 4A blue lobe has an overall P.A. of 
$\sim$160\deg\/, and its north-south appearance in Figure 1d is due to the 
relative brightness of the western cavity wall.  Figure 5 also shows dense gas 
outflowing from 4A\am\/, the binary companion to 4A, at low relative velocities 
along the south-by-southwest feature, but dense gas appears to be outflowing 
instead from 4A at higher relative velocities.

Our H$_{2}$CO data reveal a systematic decrease in the P.A. of the 4A outflow
axis with scale that could be due to precession of the 4A outflow axis (i.e., 
a systematic change in its P.A. with time.)  Single-dish CS data from B95 and 
Lefloch et al.\/ (1998) show the blue lobe oriented southwest at separations 
$>$35\as\ from 4A but oriented south at separations 15--35\as\ from 4A.  Our 
data show this trend continues to smaller separations, since the blue lobe is 
oriented to the southeast at $<$15\as\ offsets from 4A.  Such precession may 
be due to the binarity of 4A, if the gravitational action of 4A\am\ on 4A 
is strong enough (or vice-versa.)  The inherent periodicity of precession, 
however, suggests outflow gas will be distributed periodically in P.A. at even 
larger scales, and indeed such evidence may exist in recent CO maps of NGC 
1333.  For example, a CO 3--2 JCMT map of Knee \& Sandell (2000) shows what 
may be the northern red lobe of the 4A outflow curving north of IRAS 4, from a 
P.A. of 45\deg\ to a P.A. of 0\deg\ over 3--4\am.  In addition, the CO 1--0 
BIMA mosaic of NGC 1333 by Plambeck \& Engargiola (1999) faintly shows possible 
emission from the 4A outflow at large scales, curving symmetrically over 
several arcminutes from 4A in an S-shape.  However, maps that are more 
sensitive than these are required to determine conclusively if this distant 
emission is actually related to the 4A outflow.  (As this paper was being 
revised after submission, a periodicity in the P.A. of the 4A outflow was 
seen by Choi (2001).  However, the amplitude suggested here is larger by
a factor of $\sim$10.)

The difference in integrated intensity between sides of the 4A blue lobe could 
be understood if the outflow is precessing, as dense gas surrounding 4A may 
be cleared out in some directions more than others by a precessing outflow. For 
example, the emission from the western wall of the outflow cavity shown in 
Figure 5 may extend to higher relative velocities than the eastern wall if the 
outflow has encountered less mass to the west, a direction along which gas may 
have been most recently cleared out by the precessing outflow.  This scenario 
may also explain why the terminus of the outflow cavity appears to shift to the
west with increasing relative velocity.  

Figure 6 shows channel maps of the dense gas features near 4B, revealing 
the underlying structure of its outflow.  Here, all features are found at 
increasingly large positional offsets from 4B with increasing relative 
velocity in a Hubble Law manner, except the feature east of 4B which moves 
in the opposite sense.  Correspondingly, this latter feature may be emission
from dense gas outflowing from 4B\am.  (Our mosaic does not cover area east 
of 4B\am\ where other possible outflow features from 4B\am\ could reside.)

The red features to the north and southeast and the blue features to the 
south and northwest seen in Figure 6 together suggest an X-shape, possibly 
indicative of outflow cavity walls similar to those surmised for the 4A blue 
lobe.  In this case, the 4B dense gas outflow may consist of two lobes, one 
situated at a north-by-northwest P.A. (i.e., $\sim$345\deg) with an opening 
angle of $\sim$45\deg, and another at a south-by-southeast P.A. (i.e., 
$\sim$160\deg) with an opening angle of $\sim$30\deg.  The peculiar appearance 
of both red and blue outflow emission on the same sides of 4B then may be due 
to the 4B outflow being inclined from the plane-of-sky at an angle less than, 
e.g., the $\sim$30\deg\ opening angle of the narrower lobe.  An alternative 
is that the brighter and dimmer pairs of symmetric red and blue dense gas 
outflow features are each independent outflows, with a projected difference 
in the orientation of their axes of $\sim$30-45\deg, driven by unresolved 
objects within 4B.  As with 4A, the H$_{2}$CO emission may still trace cavity 
walls in this case, but here they may be unresolved.  A close companion to 
4B is suggested by the sub-arcsecond resolution continuum data of Lay et 
al.\/ and Looney et al.\/ but this has not been yet confirmed.

\bigskip
\centerline{\it 4.2.2.2 Physical Properties of Outflows}
\medskip

Physical properties of the outflowing dense gas can be estimated from the 
H$_{2}$CO data, with some assumptions.  To estimate the mass of outflowing 
gas, local thermodynamic equilibrium (LTE) was assumed and line emission was 
assumed to be optically thin, although these assumptions are likely not valid 
within dense gas outflows.  To obtain de-projected relative gas velocities, 
both outflows were assumed to originate at the $V_{LSR}$ of 4A or 4B (see \S 
4.3.2) and are inclined 10\deg\ from the plane-of-sky due to the low spatial 
coincidence of their red and blue lobes.  In addition, excitation temperature 
and H$_{2}$CO abundance values of $T_{ex}$ = 50 K and $X$(H$_{2}$CO) = 1.8 
$\times$ 10$^{-7}$ or 4.3 $\times$ 10$^{-7}$ were assumed for 4A or 4B 
respectively, as derived by B95 from wing emission.  

Re-binning the original data cube to channels of width $\Delta V$ = 0.31 km 
s$^{-1}$\/ (approximately the thermal broadening FWHM of H$_{2}$CO at 50 K, 
to obtain independent velocity channels $l$), the total gas mass of a given 
outflow feature $k$, $M_{k}$, can be estimated as

$$ M_{k} = {{Z(T_{ex})}\over{g_{j}e^{-h\nu/kT_{ex}}}}{{\mu m_{H}}\over{X({\rm H_{2}CO})}}\sum\limits_{l}^{channels}{\int\limits_{\Omega_{kl}}{{S_{\nu}^{kl}\Delta V}\over{A_{ji}h\nu}}}{D^{2}}d\Omega, \eqno{(6)}$$

\noindent
where $S_{\nu}^{kl}$ and $\Omega_{kl}$ are the average flux density and 
deconvolved area of feature $k$ in channel $l$ obtained from two-dimensional 
Gaussian fits.  (The projected distance to the nearest continuum object was 
also obtained from Gaussian fitting.)  About 6 features were identified by 
eye within each of the 4A and 4B outflows, based on their $\geq$4$\sigma$ 
peak intensities and persistence in several adjacent channels.  In addition, 
$A_{ji}$ is the Einstein A coefficient of the $3_{12}$--$2_{11}$ line, $\nu$ 
is the line frequency, $D$ is the assumed distance to IRAS 4, $Z(T_{ex})$ is 
the partition function of H$_{2}$CO at $T_{ex}$, and $g_{j}$ is the degeneracy 
of the $3_{12}$ level.

Masses ($M_{out}$) were obtained by summing Equation 6 over all features.  
Momenta ($P_{out}$) or kinetic energies ($E_{out}$) were obtained by summing 
the products of mass and relative de-projected velocity per channel, or the 
products of half the mass and the square of the relative de-projected velocity 
per channel respectively over all features.  A characteristic timescale $t$ 
was determined as the time for mass of a given feature at a given de-projected 
relative velocity to have travelled to a de-projected distance.  (Given the 
extended nature of mass in each channel, $t$ should not be mistaken for the 
actual time that all mass in the channel took to travel from the continuum 
source.)  Dividing the momenta and kinetic energies per channel by these 
timescales per channel and summing over channels and features yielded total
characteristic momentum fluxes ($F_{H_{2}CO}$, or forces) and mechanical 
luminosities ($L_{mech}$, or powers) of the outflows.

The total derived masses for the 4A and 4B dense gas outflows are 6.9 $\times$ 
10$^{-4}$ M\sun\ and 1.5 $\times$ 10$^{-4}$ M\sun\ respectively.  Table 4 lists 
these and the other physical properties derived for each dense gas outflow.  
The two brightest features associated with 4A or 4B contribute respectively 
45\% and 43\% of $M_{out}$.  Note that values for 4A listed in Table 4 include 
only the blue lobe of the 4A dense gas outflow, so total properties of the 
entire 4A dense gas outflow may be larger than those listed by a factor of 
$\sim$2, assuming the flow is symmetric, e.g., for a total $L_{mech}$ of 
0.25 L\sun\ yr$^{-1}$\/.  

Quantities in Table 4 for the 4A southern blue lobe alone exceed those 
of the all lobes of the 4B dense gas outflow by factors of 5--17.  These 
differences may be significant if both outflows have similar low inclination 
angles, but this is a very uncertain assumption.  For example, the mechanical 
luminosity, $L_{mech}$, of the 4A blue lobe would decrease by a factor of 
$\sim$90 if the inclination of the flow were 45\deg\ instead of 10\deg, due 
to the intrinsically large dependence on tan~$i$.  Differences between the 4A 
and 4B dense gas outflow characteristics could be expected from differences 
in the photon luminosities of the protostellar sources.  Millimeter wavelength
luminosities of 4A and 4B, obtained by integrating continuum flux densities 
in Table 2 and those at $\lambda$ = 450 \um\ and $\lambda$ = 850 \um\ by 
Sandell \& Knee, are 0.24 L\sun\ and 0.12 L\sun\ respectively, only a factor 
of $\sim$2 different.  Other factors, therefore, may also be relevant.

Our values of $M_{out}$, $P_{out}$, and $E_{out}$ for the 4A blue lobe are 
factors of $\sim$60--80 less than values derived for the same lobe by Knee
\& Sandell, assuming their inclination angle (45\deg) and distance (220 pc).  
These differences are likely because the CO 3--2 data of Knee \& Sandell 
(from the JCMT with a 14\as\ FWHM beam) have greater sensitivity to large 
quantities of lower density gas (i.e., $n_{cr}$[CO 3--2] $\approx$ 10$^{4}$ 
cm$^{-3}$), suggesting dense gas (as seen by the H$_{2}$CO line) is relatively 
insignficant in terms of the total mass, momentum, and energy of the molecular 
outflow, comprising $<$2\% of these quantities.  However, the momentum flux
and mechanical luminosities we derive are only factors of $\sim$10 less than 
those derived by Knee \& Sandell for the 4A blue lobe, indicating the dense 
gas has greater significance in the total force and power of the 4A outflow.  

\subsection{N$_{2}$H$^{+}$ Observations}

As discussed above, the N$_{2}$H$^{+}$ molecular ion can be relatively 
abundant in non-outflowing dense gas around protostellar objects, and the 
optical depth of its transitions can be low in all but the densest cases 
(e.g., L1544; see Williams et al.\/ 1999).  In addition, the N$_{2}$H$^{+}$ 
1--0 transition has 7 hyperfine components from which the total brightness 
temperature ($T_{b}^{tot}$), the central line velocity ($V_{LSR}$), and the 
line width ($\Delta V$) can be estimated to good accuracy via simultaneous 
hyperfine structure fitting.  These quantities can be used to estimate the 
mass of non-outflowing dense gas around 4A and 4B, and probe its rotational
and incoherent motions.  

Figure 7 shows the values of $T_{b}^{tot}$, $V_{LSR}$, and $\Delta V$ obtained
from fitting the hyperfine structure of the N$_{2}$H$^{+}$ line around 4A and 
4B, using the HFS routine of CLASS (see Caselli, Myers \& Thaddeus 1995.)   
Values of $\Delta V$ shown were obtained by deconvolving the measured $\Delta 
V$ with the 0.16 km s$^{-1}$\/ velocity resolution of the PdBI data.  Typical 
errors from HFS fitting per pixel for $T_{b}^{tot}$, $V_{LSR}$, and $\Delta V$ 
were 0.94 K, 0.015 km s$^{-1}$\/, and 0.032 km s$^{-1}$\/ respectively.  The 
quantities shown in Figure 7 are restricted to those recovered from pixels 
containing data where the peak intensity of the brightest hyperfine component 
(i.e., N$_{2}$H$^{+}$ 123-012) was $\geq$5$\times$ the rms of the respective 
spectral baseline.  With this criterion, regions of dense gas surrounding each 
object are defined, similar in extent to those obtained by the lowest contour 
of Figure 1f from an integrated intensity criterion.  

In Table 5, the mean, rms, minimum, and maximum values of $T_{b}^{tot}$, 
$V_{LSR}$, and $\Delta V$ in the non-outflowing dense gas surrounding 4A 
and 4B are summarized.  To obtain these values, the fields shown in Figure 
7 were re-binned into images with pixels 4\farcs0 $\times$ 3\farcs5 in size, 
about the area of the beam, so beam-averaged quantities at spatially
independent locations could be examined.  (Pixels not meeting the criterion 
for inclusion in Figure 7 were also not included in the re-binning.)  Around 
4B, the mean value of $T_{b}^{tot}$ is $\sim$60\% larger than it is around 
4A, but the mean values of $V_{LSR}$ and $\Delta V$ are similar in both cases.  
However, the rms of $V_{LSR}$ for the non-outflowing dense gas around 4A is 
$\sim$4$\times$ larger than that for the same gas around 4B, a point discussed
further in \S 4.3.2 below.

\subsubsection{Column Densities and Masses}

With quantities obtained from hyperfine structure fitting, the column density
of N$_{2}$H$^{+}$, $N$(N$_{2}$H$^{+}$), in a given pixel around 4A and 4B can
be estimated.  Without unique estimates of $T_{ex}$ and $\tau_{tot}$, however, 
we assume $T_{ex}$ = 10 K $>>$ $hB/3k$, $T_{bg}$ = 2.73 K.\footnote{$T_{bg}$ 
is much larger at the positions coincident with the peak continuum intensity, 
but since this is only significant for relatively few pixels, that difference 
is ignored here.}.  In addition, we assume the line is optically thin (i.e., 
$T_{b}^{tot}$ $\approx$ $T_{ex}\tau_{tot}$), which is supported by the 
symmetric appearance of the line over the regions shown in Figure 7.  Following 
similar estimates by Womack et al.\/, but with these assumptions,

$$ N({\rm N_{2}H^{+}}) \approx {10^{5}\over{1.06(J+1)}}{3h\over{8\pi^{3}\mu^{2}}}{k(T_{b}^{tot})\over{hB}}{e^{E_{J}/kT_{ex}}\over{(1-e^{-h\nu/kT_{ex}})}}{\Delta V}, \eqno{(7)} $$

\noindent
where $J$ is the upper rotational level (1), $\mu$ is the dipole moment of the
N$_{2}$H$^{+}$ molecular ion (3.40 Debye; Green, Montgomery \& Thaddeus 1974),
$B$ is the rotational constant of N$_{2}$H$^{+}$ (46.586702 GHz; Caselli et 
al.\/), $E_{J}$ is the upper level energy (= $h\nu$ where $\nu$=93.1734035 GHz; 
Caselli et al.), and $T_{b}^{tot}$, and $\Delta V$ come from HFS fits.  (Note 
that units for $\Delta V$ are km s$^{-1}$\/.)

From Equation 7, mean values of $N$(N$_{2}$H$^{+}$) in the non-outflowing 
dense gas around 4A and 4B of 2.4 $\times$ 10$^{12}$ cm$^{-2}$ and 3.9 $\times$ 
10$^{12}$ cm$^{-2}$ are found respectively.  Table 5 also lists the rms, 
minimum, and maximum values for $N$(N$_{2}$H$^{+}$) obtained for this gas 
around each protostellar object.  (Decreasing $T_{ex}$ to 5 K reduces the 
values of $N$(N$_{2}$H$^{+}$) in Table 5 by only factors of $\sim$5\%, but 
increasing $T_{ex}$ to 20 K increases them by factors of $\sim$40-50\%.)  The 
mean column densities of non-outflowing dense gas around both objects are 
similar, although that of the gas around 4B is $\sim$60\% larger than that 
around 4A.  These values are 2.5-5$\times$ larger than the $N$(N$_{2}$H$^{+}$) 
of 0.95 $\times$ 10$^{12}$ cm$^{-2}$ obtained from the mean N$_{2}$H$^{+}$ 
spectrum of IRAS 4 from single-dish data of M98, under the same assumptions.  
Maximum column densities listed in Table 5 are 5-8$\times$ larger than this 
latter value.  (The $N$(N$_{2}$H$^{+}$) quantities listed in Table 5 are 
actually lower limits, given the assumption of $\tau_{tot}$ $\rightarrow$ 0, 
but the quantities should not vary much from those listed if $\tau_{tot}$ 
$<$ 1 as expected.)

With $N$(N$_{2}$H$^{+}$), the masses of non-outflowing dense gas traced by 
N$_{2}$H$^{+}$\/, $M_{gas}$\/, can be estimated by assuming the emission 
originates from gas local to 4A or 4B and not from gas distributed along 
the line-of-sight.  With this assumption, 

$$ M_{gas} = {\mu m_{H}\over{X(\rm{N_{2}H^{+}})}} D^{2} \int_{\Omega_{s}} N({\rm N_{2}H^{+}})d\Omega, \eqno{(8)} $$

\noindent
where $\mu$, $m_{H}$, and $D$ are the same as for Equation 6, but $\Omega_{s}$ 
is the solid angle subtended by the dense gas traced by N$_{2}$H$^{+}$ around 
each object, and $X({\rm N_{2}H^{+}})$ is the fractional abundance of 
N$_{2}$H$^{+}$.

For $X({\rm N_{2}H^{+}})$, we assume for 4A and 4B the value of 6.0 $\times$ 
10$^{-11}$ derived by B95 at 4A by comparing the observed column density of 
N$_{2}$H$^{+}$ with that suggested for H$_{2}$ by continuum observations of 
dust.  This value is relatively low compared to mean values found in other 
cloud cores; Womack et al.\/ and Benson et al.\/ found mean values of $X({\rm 
N_{2}H^{+}})$ of 4 $\times$ 10$^{-10}$ and 7 $\times$ 10$^{-10}$ respectively 
from various samples (see also Ungerechts et al. 1997).  However, the sample 
of Womack et al. contains several cases where $X({\rm N_{2}H^{+}})$ is as low 
as or lower than 6.0 $\times$ 10$^{-11}$, e.g., $\rho$ Oph and W49.  Assuming 
a ``canonical" CO abundance of 1 $\times$ 10$^{-4}$, B95 also derived a value 
for $X({\rm N_{2}H^{+}})$ of 6.0 $\times$ 10$^{-10}$, more similar to the mean
values of other cloud cores, by comparing column densities of N$_{2}$H$^{+}$ 
with that of H$_{2}$ suggested by C$^{18}$O and C$^{17}$O data.  B95 suggested 
this disparity could be reconciled if CO (and other molecules) in the 
non-outflowing gas of IRAS 4 were depleted relative to dust by factors of 
10--20, but maintained abundances relative to each other that were similar to 
those seen elsewhere.  

Using Equation 8 and the above assumptions, the $M_{gas}$ of 4A and 4B are 
0.73 M\sun\ and 0.41 M\sun\ respectively.  As a comparison, recent SCUBA 
observations of continuum emission from dust around 4A and 4B by Sandell \& 
Knee yield circumstellar mass estimates of 2.0-4.0 M\sun\ and 0.65-1.3 M\sun, 
following their analysis but assuming a distance to NGC 1333 of 350 pc and 
a range of $T_{K}$ of 20-40 K.  (In their analysis, Sandell \& Knee assumed 
$T_{K}$ = 25 K and a 220 pc distance to derive masses.)  The SCUBA-based 
masses are somewhat larger than ours, possibly due to their greater sensitivity 
to larger-scale emission from lower-density material.  

Another comparison of mass is provided by virial masses of the non-outflowing
dense gas surrounding 4A and 4B.  These can be calculated from the respective 
mean values of $\Delta V$ and apparent sizes as 

$$ M_{vir} = 210 \Biggl({{R}\over{{\rm pc}} }\Biggr) \Biggl({ {\Delta V}\over{{\rm km\ s^{-1}}} }\Biggr)^{2} M_{\odot}, \eqno{(9)} $$

\noindent
for a uniform, self-gravitating sphere (e.g., see Larson 1981).  Using the 
mean values of $\Delta V$ listed in Table 4, and the values of $R$ of 0.022 
pc for 4A and 0.0093 pc for 4B (obtained from the extents of the integrated 
emission shown in Figure 1f) in Equation 9, virial masses of 1.1 M\sun\ and 
0.48 M\sun\ are found for the non-outflowing dense gas around 4A and 4B 
respectively, values quite similar to those determined above from the derived 
column densites.  (Alternatively, values of $X({\rm N_{2}H^{+}})$ of 4.0 
$\times$ 10$^{-11}$ and 5.1 $\times$ 10$^{-11}$, similar to that found by B95 
for 4A, can be found for 4A and 4B respectively by inverting Equation 8 and 
assuming $M_{vir}$ = $M_{gas}$.) 

\subsubsection{Rotation}

Figures 7b and 7e show the distributions of $V_{LSR}$ with position around 
4A and 4B respectively obtained from hyperfine structure fitting.  Systematic 
variations with position of the central velocity, $V_{LSR}$, of the symmetric 
profiles of an optically-thin line can suggest velocity gradients in the gas.  
If a gradient is found symmetrically about an average velocity and position, 
it may be caused by rotational motions of the gas.  

As listed in Table 5, the mean V$_{LSR}$ values of the non-outflowing dense 
gas associated with 4A and 4B are 6.95 km s$^{-1}$\/ and 6.98 km s$^{-1}$\/ 
respectively, each within the mean fitting uncertainty of the other (0.032 km 
s$^{-1}$\/).  If these mean velocities of this dense gas are representative 
of the radial velocities of the 4A and 4B objects themselves, the 4A and 
4B objects are not moving relative to each other significantly along the 
line-of-sight, as might be expected if 4A and 4B were revolving about each 
other.  (However, a $V_{LSR}$ difference of 0.28 km s$^{-1}$\/ is found 
between 4A and 4B from modeling their respective associated inverse P-Cygni 
profiles; see \S 4.2.1, Table 3.)  The average $V_{LSR}$ obtained from the 
non-outflowing dense gas around both objects is 6.96 km s$^{-1}$, which we 
take as the systemic velocity of IRAS 4.

Near 4A, the rms variation of $V_{LSR}$ is more than half the mean $\Delta V$, 
i.e., 0.36 km s$^{-1}$\/ vs. 0.25 km s$^{-1}$\/, suggesting the variations of 
$V_{LSR}$ around 4A are significant.  Furthermore, the positional distribution 
of $V_{LSR}$ in the non-outflowing dense gas about 4A shows systematic, 
symmetrical variations that may due to rotation.  To first order, the 
southeastern half of this gas around 4A is blue-shifted (relative to the 
mean) by $\sim$0.25-0.5 km s$^{-1}$\/ and the northwestern half is red-shifted 
by $\sim$0.0-0.5 km s$^{-1}$\/, variations that could arise if this gas were 
rotating about 4A on an axis at a P.A. of $\sim$45\deg.  (Such an axis would 
be colinear with the ``cavity" containing 4A and 4A\am.)  Rotating dense gas, 
however, cannot account for all variations of $V_{LSR}$ about 4A.  For example, 
two pockets to the north-by-northwest and one to the south-by-southwest have 
relative velocities in directions opposite to those expected from rotation 
at their locations.  Given that the south-by-southwest pocket is unusually 
red-shifted and that it lies in the same direction of the blue 4A outflow 
lobe (see \S 4.2.2), the pocket may be a local distortion of a rotational 
velocity field due to selective clearing of N$_{2}$H$^{+}$ at blue-shifted 
velocities by the outflow in that direction.  A similar explanation may 
account for the unusually blue-shifted pockets in the same direction as the 
red outflow lobe.  (The interpretation of velocity variations as caused by 
rotation is not unique, and the velocity distribution also could be due to 
bulk infall or outflow motions of gas on either side of 4A.)

Near 4B, the characteristics of $V_{LSR}$ are much different from those seen
around 4A.  The rms variation of $V_{LSR}$ is less than half the mean $\Delta 
V$, i.e., 0.096 km s$^{-1}$\/ vs. 0.25 km s$^{-1}$\/, and only $\sim$25\% 
that of the non-outflowing dense gas around 4A, suggesting the variations 
of $V_{LSR}$ are not significant.  In addition, no systematic variation of 
$V_{LSR}$ is seen with position, and hence no suggestion of rotation (and 
infall or outflow), at least from the N$_{2}$H$^{+}$ 1--0 line.  The 
difference in rotational properties between the non-outflowing dense gas 
surrounding 4A and 4B is striking in the context of binary formation.  
Since 4A is a known binary of 1\as\ (or 350 AU) separation but 4B has not 
been confirmed to be a multiple, the relatively large rotation of the 
non-outflowing dense gas about 4A suggests binaries may form preferentially 
in cores of relatively high intrinsic angular momentum but single stars form 
preferentially in cores of relatively low intrinsic angular momentum.  
(However, 4B may be yet a binary system, albeit with a much smaller separation 
than that of 4A and 4A\am; see \S 4.2.2). 

\subsubsection{Turbulent Motions}

NGC 1333 is located in the western Perseus molecular cloud, a region where 
a large number of stars are forming in close proximity from gas that is fairly
turbulent, judging from the observed line widths.  For example, IRAS 4A and 4B 
have among the largest N$_{2}$H$^{+}$ 1--0 line widths, 1.32 km s$^{-1}$\/ 
and 1.04 km s$^{-1}$\/ respectively, of the 47 Class 0 or Class I objects 
within 400 pc sampled by M97.  As the thermal FWHM of N$_{2}$H$^{+}$ is 0.247 
km s$^{-1}$\/ if $T_{K}$ = 40 K (i.e,. the maximum of the $T_{K}$ range derived 
by B95 for the non-outflowing gas), the broad lines seen by M97 in IRAS 4 
suggest the presence of substantial turbulent motions in the dense gas on 
scales of 0.05 pc (i.e., the 27\as\ IRAM 30-m Telescope beam FWHM at 93.17 
GHz at 350 pc.)

The PdBI data described here probes physical scales much smaller than those 
of single-dish observations, e.g., the synthesized beam FWHM is only 0.0055 
$\times$ 0.0064 pc at 350 pc.  Figures 7c and 7f show the distributions of 
the N$_{2}$H$^{+}$ 1--0 line width near 4A and 4B respectively, i.e., values 
of $\Delta V$ obtained from hyperfine structure fitting.  Table 5 lists the 
beam-averaged mean, rms, maximum and miniumum values of $\Delta V$ found 
for non-outflowing dense gas around 4A or 4B.  From these data, a dramatic 
reduction in N$_{2}$H$^{+}$ line width is seen, relative to the single-dish 
data.  For example, maximum beam-averaged values of $\Delta V$ around both 
objects are $<$0.84 km s$^{-1}$\/, and the mean beam-averaged values of $\Delta 
V$ are both only $\sim$0.50 km s$^{-1}$\/, much less than seen in single-dish 
data.  

At first glance, it is conceivable that the larger $\Delta V$ in single-dish
observations could originate from the unresolved combination of intrinsically 
narrow lines with potentially wide variations of central line velocity in the 
dense gas (e.g., from rotation).  However, the rms values of $V_{LSR}$ and the
mean values of $\Delta V$ are too small to support this possibility.  Instead, 
the overall reduction of line width with scale in IRAS 4 may be a manifestation 
of the linewidth-size relation, as observed in many molecular clouds (e.g., see 
McKee 1999 and references therein), and the interferometer data is probing the 
smallest scales of this relation.  At these scales, the nonthermal component 
of the line width is predicted to be comparable in magnitude to the thermal 
component, possibly due to dissipation of turbulence (Vazquez-Semadeni et al.\/ 
1999) or the cutoff of MHD wave propagation (Nakano 1998; Myers 1998) at small 
scales.

A similar but less dramatic reduction of line width with scale was noted for 
8 ``quiescent cores" in the Serpens NW cluster region by Williams \& Myers, 
which is presumed to be at a similar distance to NGC 1333 (i.e., 310 pc; de 
Lara, Chavarria-K, \& Lopez-Molina 1991).  In Serpens NW, the mean width ratio 
for lines observed between 10\as\ and 50\as\ resolution over all 8 cores was 
0.59, whereas the mean width ratio for lines observed between $\sim$4\as\ and 
27\as\ over 4A and 4B is 0.43, using the observations of M98 and the mean 
values of $\Delta V$ in Table 5.  Interestingly, the 8 cores found by Williams 
\& Myers have typical sizes of $\sim$5000 AU, line widths of $\sim$0.5 km 
s$^{-1}$\/, and infall speeds of $\sim$0.5 km s$^{-1}$\/, values all similar 
to those of the non-outflowing dense gas around 4A and 4B (assuming the infall 
speeds found in \S 4.2.1 from H$_{2}$CO apply to the N$_{2}$H$^{+}$ gas.)

Despite the overall relative reduction in line width with scale seen here, the 
high spatial resolution of these observations reveals significant variations in 
$\Delta V$ in the non-outflowing dense gas around both objects (see Figures 7c 
and 7f), and a common narrow width of $\sim$0.5 km s$^{-1}$\/ cannot describe 
the line emission.  Figure 8 shows histograms of the distributions of $\Delta 
V$ as fractions of the total number of beam-averaged pixels around 4A (54 
pixels or 4B (12 pixels) in 0.15 km s$^{-1}$ bins.  Both have ranges of $\Delta 
V$ between 0.23 km s$^{-1}$\/ to 0.83 km s$^{-1}$\/, with substantial fractions (15-30\%) near 0.25 km s$^{-1}$, the thermal line width at 40 K, indicating the 
presence of very non-turbulent gas around each object.  (Minimum beam-averaged 
$\Delta V$ values of 0.21 km s$^{-1}$\/ near 4A and 0.16 km s$^{-1}$\/ near 4B 
are found, located north of 4A and 4B, or at the peripheries of the dense gas.) 
Of the remainder, the distribution for 4A has peaks at 0.38 km s$^{-1}$\/ and 
0.68 km s$^{-1}$\/, but that of 4B has only a peak at 0.23 km s$^{-1}$\/.  For
4A, lines of the first peak are located generally to the northwest and lines of 
the second peak are located generally to the southeast, although the populations
are mixed.  For 4B, the lines broaden systematically from north to south across 
its dense gas.

The condensations from which 4A and 4B formed may be plausibly assumed to 
have had nearly-uniform initial velocity dispersions, as is seen in many 
starless dense cores (see Goodman et al.\/ 1998; Caselli et al.\/ 2001).  The 
star formation process from that time until the present likely increased the 
velocity dispersion by differing amounts thoughout each condensation, due 
mainly to modest outflow-core interaction or heating from the protostars, 
but very likely did not decrease the dispersion anywhere.  Every independent 
measurement of line width in the gas around each object then gives an upper 
limit on the initial velocity dispersion of the original condensation.  The 
typical upper limit comes from the mean present-day line width, $\sim$0.5 
km s$^{-1}$\/ in both 4A and 4B, and the most constraining upper limit 
comes from the narrowest present-day line width, $\sim$0.2 km s$^{-1}$\/ in 
the gas around both 4A and 4B.  These new limits indicate that the initial 
level of turbulence in star-forming condensations may be very low, even in 
regions of clustered star formation.  For example, if $T_{K}$ = 20 K (the 
low end of $T_{K}$ from B95) and $\Delta V$(observed) = 0.225 km s$^{-1}$\/ 
(the lowest $\Delta V$ bin in Figure 8), then $\Delta V$(thermal) = 0.18 km 
s$^{-1}$\/ for N$_{2}$H$^{+}$ and $\Delta V$(nonthermal) = 0.135 km s$^{-1}$\/, 
as $\Delta V$(nonthermal) = ($\Delta V^{2}$(observed) - $\Delta V^{2}$(thermal))
$^{0.5}$.  

If the velocity dispersion has evolved from its initial state though the 
injection of energy, line widths arguably should be largest at positions 
closest to possible sources of energy, and there is indeed some evidence 
for this in the dense gas around both objects.  For example, there appears 
to be some spatial correspondence between large values of $\Delta V$ and 
the blue-shifted outflow features (as traced by H$_{2}$CO) south of both 
4A and 4B.  However, the situation is not so clear cut and evidence to 
the contrary is also found.  For example, the narrowest lines in the 
non-outflowing dense gas around 4A and 4B are found actually near the 
respective red outflow features north of each object.  Moreover, large 
line widths are found across the southeastern half of the 4A dense gas, 
not just at positions near the blue outflow lobe.  Given the factor of 
$\sim$10--20 difference between the spectral width of the H$_{2}$CO wings 
and the width of the N$_{2}$H$^{+}$ line even at positions of large $\Delta 
V$, whatever coupling does exist between the dense outflowing gas and 
N$_{2}$H$^{+}$ appears to be very inefficient.  

The weak coupling between the fast outflow and the non-outflowing gas 
may be a consequence of their difference in speed, with greater coupling 
expected when the flow speed is more comparable to the velocity dispersion 
of its environment.  On the other hand, possibly the N$_{2}$H$^{+}$ is depleted 
more rapidly than it can be accelerated to speeds similar to the outflow.  
(The contrast of widths cannot be ascribed to the outflow gas being less 
dense than the non-outflowing gas, as both transitions probe gas at $n$ 
$\approx$ 10$^{5-6}$ cm$^{-3}$.)  Similar contrasts in line width have been 
found toward other very young objects, but the proximity of outflow gas to 
the non-outflowing gas is not so clearly defined in single-dish observations.
For example, the survey of M97 contains several examples with outflow wings 
in H$_{2}$CO and CS, but not in N$_{2}$H$^{+}$, on 0.02--0.05 pc scales, 
toward sources such as L483, S68N, or L1157 (and, of course, 4A and 4B).  
The PdBI data here show emission from both outflowing and non-outflowing gas 
around 4A and 4B can be spatially coincident on scales of only $\sim$0.006 
pc.

\section{Summary and Conclusions}

Millimeter interferometer observations are presented of NGC 1333 IRAS 4, a 
group of highly-embedded protostellar objects in the Perseus star-forming 
complex.  Millimeter continuum data and line data allow the immediate dense 
gas and dust environments of its protostellar objects to be characterized 
simultaneously at high spatial resolution.  The major results found from 
these observations include:

1. Inverse P-Cygni profiles of H$_{2}$CO 3$_{12}$--2$_{11}$ that show 
red-shifted absorption and blue-shifted emission have been detected toward 
4A and 4B.  Inverse P-Cygni profiles of CS 3--2 and N$_{2}$H$^{+}$ 1--0 also
have been detected toward 4A.  These profiles provide the least ambiguous 
evidence for infalling gas motions associated with low-mass young stellar 
objects.  The depth of the absorption in the H$_{2}$CO line are equal to the 
$\lambda$ = 1.3 mm continuum intensities of the protostars along the same 
lines-of-sight.  Modeling the H$_{2}$CO profiles with a ``two-layer" radiative 
transfer code, infall velocities toward 4A and 4B of 0.68 km s$^{-1}$\/ and 
0.47 km s$^{-1}$\/ are estimated respectively.  From these, masses of 0.53 
M\sun\ and 0.17 M\sun, and accretion rates, $\dot M_{in}$, of 1.1 $\times$ 
10$^{-4}$ M\sun\ yr$^{-1}$\/ and 3.7 $\times$ 10$^{-5}$ M\sun\ yr$^{-1}$\/ 
for 4A and 4B are estimated respectively, given assumptions abut $r_{in}$ 
and $n$.  Accretion timescales for both objects are similar, with $t_{acc}$ 
= $\dot M$/$M$ of 4700 yr and 4600 yr for 4A and 4B respectively.

2. Outflowing gas from both 4A and 4B is detected from bright wing emission 
of H$_{2}$CO 3$_{12}$--2$_{11}$ and CS 3--2.  The outflows from each source 
differ, in that only the southern blue lobe from the extended outflow from 
4A is detected but 4-5 compact features from 4B are detected.  For 4A, the 
velocity structure of the emission suggests the H$_{2}$CO is tracing outflow 
cavity walls.  The orientation of the cavity walls suggests the outflow has 
precessed from orientations evident at larger scales, due possibly to the 
binarity of 4A.  For 4B, the multipolar structure could be also from cavity 
walls of an outflow that is more compact than that from 4A.  However, the 
complex structure may originate from several juxtaposed lobes from a multiple 
object.  

3. Non-outflowing dense gas surrounding 4A and 4B individually is detected 
from N$_{2}$H$^{+}$ 1--0 emission.  With assumptions, we determine mean 
column densities of 2.4 $\times$ 10$^{12}$ cm$^{-2}$ and 3.9 $\times$ 10$^{12}$ 
cm$^{-2}$ respectively.  Furthermore, masses of this gas are estimated to be 
0.73 M\sun\ and 0.41 M\sun\ respectively, comparable to other estimates of 
circumstellar mass derived from submillimeter dust emission and the virial 
theorem.  The variations of $V_{LSR}$ around each object suggest the 
non-outflowing dense gas about 4A is rotating, but the data do not suggest 
similar motions about 4B.

4. The widths of the N$_{2}$H$^{+}$ 1--0 line in the non-outflowing dense 
gas around both 4A and 4B are smaller by factors of 2--3 on average than those 
observed in single-dish data, due to the reduction of turbulent motions on 
smaller physical size scales.  The data reveal a new, low upper bound to the 
initial velocity dispersion in the protostellar core from which IRAS 4 formed.  
For example, some regions around each object ($\sim$15-30\% of the projected 
surface areas of line emission) exhibit lines with very narrow widths, e.g., 
$\sim$0.2 km s$^{-1}$\/.  Such narrow line widths are consistent with the 
incoherent motions of the gas having a nonthermal component of a magnitude 
comparable to that of thermal broadening at expected gas temperatures, and 
may represent the initial state of the core gas prior to the formation of the 
protostars or their outflows.  The lines show considerable variation of width, 
however, even on the small scales probed.  Coupling between outflow and the 
non-outflowing dense gas, that could inject turbulent energy into any initially 
non-turbulent gas, appears highly inefficient, as line widths from H$_{2}$CO 
and N$_{2}$H$^{+}$ can differ by factors of 10--20 on scales of only 0.006 pc.

\acknowledgments 
We acknowledge the generosity of many people who assisted with this study, 
in particular A. Dutrey, R. Neri, and the staff of the IRAM Plateau de Bure 
Observatory in France, M. Momose, N. Okumura, and the staff of the Nobeyama 
Radio Obsevatory in Japan, and M. Rupen, and the staff of the NRAO AOC in 
Socorro, New Mexico, U.S.A.  We also thank an anonymous referee for many 
helpful comments, as well as Neal Evans, Lee Mundy, and Jeff Mangum for 
useful discussions.  JD thanks the Smithsonian Astrophysical Observatory for 
a generous Fellowship, during which this study was begun.  PCM acknowledges 
support from NASA Origins grant NAG5-6266.  DM acknowledges support from 
FONDECYT 1990632.

\appendix

\section{High-Resolution Continuum Observations of NGC 1333 IRAS 4}

Table 2 lists the flux densities, peak intensities, and deconvolved sizes of 
each source, derived from fitting single-component two-dimensional Gaussians 
to observations of each source individually at each wavelength with the AIPS
task IMFIT.  (Positions are also defined by these fits, e.g., those shown 
in Figure 1 are those of the peak continuum intensties at $\lambda$ = 1.3 mm
from Gaussian fitting.)  Flux density uncertainties are comprised of those 
derived from the fits, an assumed 15\% uncertainty in the absolute calibration 
at each wavelength, and an estimated 10\% measurement uncertainty.  Values in 
Table 2 for 4B\am\ at $\lambda$ = 1.3 mm are particularly uncertain due to 
its location at the mosaic edge (see Figure 1a), although the peak position 
obtained from the fit agrees quite well with those from other wavelengths.

Figure A1 shows the millimeter wavelength spectral energy distributions 
(SEDs) of each IRAS 4 object from our continuum data.  In addition, Figure 
A1 shows data obtained at $\lambda$ = 800 $\mu$m by Lay et al.\/ from the 
JCMT-CSO Interferometer, at $\lambda$ = 1.3 cm and $\lambda$ = 3.6 cm by
Mundy et al.\/ (1993) at the VLA, and at $\lambda$ = 3.6 cm and $\lambda$ 
= 6.0 cm by Rodriguez et al.\/ (1999) at the VLA.  At all wavelengths, 4A 
is brighter than 4B, which is in turn brighter than 4B\am.  The SEDs of 
Figure A1 are reasonably well fitted by straight lines of similar slopes, 
i.e.,. $\alpha_{1}$ = -2.7 $\pm$ 0.05 for 4A, -2.9 $\pm$ 0.14 for 4B, and 
-2.6 $\pm$ 0.23 for 4B\am, using linear least-squares fitting over the range 
1.3 mm $<$ $\lambda$ $<$ 1.3 cm.  (For 4A and 4B, only our data at $\lambda$ 
= 1.3 cm was included for the fit, as they are similar (i.e., within errors) to 
those found by Mundy et al.\/ at the same wavelength.)  For 4B\am, the linear 
fit included the flux density at $\lambda$ = 1.3 mm, although this measurement 
is particularly uncertain, and excluded our upper limit at $\lambda$ = 1.3 cm.

The similar slopes reveal that continuum emission for the IRAS 4 objects 
at $\lambda$ $<$ 1.3 cm is dominated by thermal emission from dust grains 
surrounding each source.  Given the deconvolved sizes of the emission, the 
warm dust being traced is likely situated within accretion disks or the 
inner radii of envelopes around each source.  If the dust emission is 
optically thin (and we assume the same region is probed) at all wavelengths, 
the slope, $\alpha$, can be used to estimate the indices of the mass opacity 
of dust, i.e., $\beta$\/.  As -$\alpha$ = $\beta$ + 2 in the Rayleigh-Jeans 
limit, dust opacity indices of 0.7 $\pm$ 0.05 for 4A, 0.9 $\pm$ 0.14 for 4B, 
and 0.6 $\pm$ 0.25 for 4B\am\ are found from our $\alpha_{1}$, each consistent 
with a common value of $\beta$ $\approx$ 0.75, and significantly less than 
the $\beta$ = 2 value assumed typically for dust in the ISM.  These values 
are similar to the $\beta$ $\approx$ 1 values inferred for dust around many 
other young stellar objects, a decrease in $\beta$ possibly due to growth 
or composition changes of the dust (Beckwith \& Sargent 1991).

The submillimeter wavelength data of Lay et al.\/, not included in the linear 
fits shown in Figure A1, deviate from the derived fits because it was obtained 
at a resolution of $\sim$1\as, and much of the emission from dust in the inner 
radii of envelopes may have been resolved out in those data.  The centimeter 
wavelength data of Rodriguez et al.\/, also not included in the linear fits, 
deviate dramatically from the lines fit solely to the data at $\lambda$ $\leq$ 
1.3 cm, suggesting that free-free emission becomes increasingly dominant in the 
SEDs of these sources at long wavelengths.  Extrapolating these data to shorter 
wavelengths, the contribution of this emission to the respective flux densities 
is small, i.e., $<$20\% at $\lambda$ = 1.3 cm and $<$5\% at $\lambda$ = 6.9 mm 
for both 4A or 4B.

Departures of individual flux densities at millimeter wavelengths from the 
linear fits are seen least in the fit to the 4B SED, whose reduced $\chi^{2}$ 
= 1.6, but are seen more for the fits to the 4A SED (reduced $\chi^{2}$ = 
7.9) and 4B\am\ SED (reduced $\chi^{2}$ = 9.5).  These latter large values 
of reduced $\chi^{2}$ may be due to underestimates of the absolute flux 
calibration uncertainties.

At specific wavelengths, the largest departures from the linear fits are the 
positive ones found at $\lambda$ = 3.2 mm for all 3 objects.  The object 
3C~84 was also observed during all tracks as IRAS 4, but a similar positive 
departure of its flux density at $\lambda$ = 3.2 from a linear fit is not seen. 
This consistency suggests the deviations in the IRAS 4 SEDs from their fits 
at this wavelength may be real, and the slopes of the IRAS 4 SEDs at $\lambda$ 
$\leq$ 3.2 mm may be shallower than those found in earlier fits.  Dust emission 
at $\lambda$ $>$ 3.2 mm may be declining at a steeper overall slope, possibly 
because the cooler, outer layers of the IRAS 4 envelopes are resolved out at 
our high resolutions, or have lower optical depths than the inner regions.

Slightly shallower slopes of the linear fits are found for the 4A and 4B SEDs 
if our VLA data are additionally excluded from the fits, i.e., $\alpha_{2}$ = 
-2.6 $\pm$ 0.29 and -2.7 $\pm$ 0.30 respectively, but these do not differ 
significantly from slopes obtained with the VLA data included.  One slope that 
does differ significantly is that of 4B\am, namely $\alpha_{2}$ = -1.8 $\pm$ 
0.39.  The slopes of all three SEDs change only slightly if the NMA data are 
also excluded.  Unlike for 4A and 4B, both $\alpha_{1}$ and $\alpha_{2}$ for
4B\am\ are significantly more than the values of -3.8 or -3.9 derived by Choi 
(2001) from interferometer continuum data at $\lambda$ = 3.4 mm and $\lambda$ 
= 2.7 mm.  Values of $\beta$ derived from our slopes suggest so 4B\am\ could 
have dust with significantly different opacity than 4A and 4B, possibly from 
a population of larger dust grains.  Alternatively, the dust may be optically 
thick.  Given the relative dimness of 4B\am\ relative to 4A or 4B, more 
sensitive data than those shown here (especially at $\lambda$ = 1.3 mm and 
$\lambda$ = 6.9 mm) are needed to properly evaluate either possibility.

\clearpage

\begin{table*}
\caption{Log of Observations} \label{tbl-1}

\begin{center}
\begin{tabular}{ccccccc}
\tableline
$\lambda$ & $\nu$ & Interferometer & Configuration & Observing Dates & Beam FWHM & Continuum rms\cr  
(mm) & (GHz) & & & & (\as) & (mJy beam$^{-1}$)\cr
\tableline
\tableline
\\
1.3 & 225.7 & IRAM PdBI &      5C2-W09 & 21 Jul 1997 & 2.0$\times$1.7 & 17.5\\  
    &       &           & 4ant-special & 19 Aug 1997\\
    &       &           &       5D-N09 & 19 Sep 1997\\
    &       &           &      4D2+N13 & 14 Nov 1997\\
\\
2.0 & 147.0 &       NMA &            D & 27 Dec 1997 & 3.2$\times$2.3\tablenotemark{a} & 14.0\\  
    &       &           &            C & 09 Mar 1998\\
\\
2.2 & 137.0 &       NMA &            D & 27 Dec 1997 & 3.4$\times$2.6 & 10.5\\  
    &       &           &            C & 09 Mar 1998\\
\\
3.2 &  93.2 & IRAM PdBI &      5C2-W09 & 20 Jul 1997 & 2.9$\times$2.6 &  4.5\\  
    &       &           &      5C2-W09 & 21 Jul 1997\\
    &       &           & 4ant-special & 19 Aug 1997\\
    &       &           &       5D-N09 & 19 Sep 1997\\
    &       &           &      4D2+N13 & 14 Nov 1997\\
\\
6.9 &  43.3 &       VLA &            D & 25 Nov 1997 & 2.1$\times$1.9 &  0.30\\  
\\
13.3&  22.5 &       VLA &            D & 25 Nov 1997 & 3.8$\times$3.1 &  0.28\\  
\\
\tableline
\end{tabular}
\end{center}

\tablenotetext{a}{CS 3--2 data tapered to a 6\farcs5 $\times$ 4\farcs4 
FWHM beam for Figure 1e and analysis.}

\end{table*}

\clearpage

\begin{table*}
\caption{Integrated Flux Densities, Peak Intensities and Deconvolved Sizes} \label{tbl-2}

\begin{center}
\begin{tabular}{cccccc}
\tableline
       &           &                   &                  & \multicolumn{2}{c}{Deconvolved Size} \cr
Object & $\lambda$ & $S_{\nu}^{total}$ & $I_{\nu}^{peak}$ & Major $\times$ Minor Axis & P.A. \cr
       &    (mm)   &        (mJy)       & (mJy beam$^{-1}$) & (AU $\times$ AU) & (degrees)\cr
\tableline
\tableline
 \\
 4A &   1.3 & 3100 $\pm$ 470 & 1200 $\pm$ 16. & 920 $\times$ 720 & 132 \\
    &   2.0 &  690 $\pm$ 110 & 450 $\pm$ 14. & 770 $\times$ 590 & 129 \\
    &   2.2 &  580 $\pm$ 90. & 420 $\pm$ 10. & 740 $\times$ 570 & 119 \\
    &   3.2 &  320 $\pm$ 49. & 200 $\pm$ 4.4 & 790 $\times$ 650 & 177 \\
    &   6.9 &  26. $\pm$ 4.0 & 16. $\pm$ 0.29 & 580 $\times$ 490 & 138 \\
    &  13.3 &  4.7 $\pm$ 0.95 & 3.1 $\pm$ 0.27 & 1100 $\times$ 630 & 125 \\
 \\
\tableline
 \\
 4B &   1.3 & 1100 $\pm$ 170 & 590 $\pm$ 17. & 600 $\times$ 560 & 151 \\
    &   2.0 &  320 $\pm$ 56. & 250 $\pm$ 14. & 630 $\times$ 390 & 119 \\
    &   2.2 &  240 $\pm$ 41. & 210 $\pm$ 11. & 440 $\times$ 300 & 151 \\
    &   3.2 &  93. $\pm$ 16. & 90. $\pm$ 4.5 & $<$570 $\times$ $<$570 & ...\\
    &   6.9 &  7.8 $\pm$ 1.3 & 5.8 $\pm$ 0.29 & 630 $\times$ 110 & 128 \\
    &  13.3 &  1.2 $\pm$ 0.56 & 1.0 $\pm$ 0.28 & $<$1400 $\times$ $<$1400 & ...\\
 \\
\tableline
 \\
4B\am &1.3 &  230 $\pm$ 47. & 200 $\pm$ 18. & $<$640 $\times$ $<$640 & ...\\
   &   2.0 &  79. $\pm$ 29. & 70. $\pm$ 14. & $<$840 $\times$ $<$840 & ...\\
   &   2.2 &  82. $\pm$ 21. & 89. $\pm$ 10. & $<$780 $\times$ $<$780 & ...\\
   &   3.2 &  45. $\pm$ 11. & 41. $\pm$ 4.5 & $<$690 $\times$ $<$690 & ...\\
   &   6.9 &  2.2 $\pm$ 0.65 & 1.9 $\pm$ 0.30 & $<$840 $\times$ $<$840 & ...\\
   &  13.3 & $<$0.84 & $<$0.84 & ... & ... \\
 \\
\tableline
\end{tabular}
\end{center}
\end{table*}

\clearpage

\begin{table*}
\caption{``Two-layer" Model Parameters for Inverse P-Cygni Profiles\tablenotemark{a}}\label{tbl-3}

\begin{center}
\begin{tabular}{ccccccccc}
\tableline
Object & $\tau_{o}$ & $\Phi$ & $J_{c}$ & $J_{f}$ & $J_{r}$ & $V_{LSR}$  & $\sigma$ & $V_{in}$ \\ 
       &            &        &    (K)     &     (K)     &     (K)    &  (km s$^{-1}$)   &    (km s$^{-1}$)   & (km s$^{-1}$) \\ 
\tableline
\tableline
 \\
4A & 2.0 & 0.3 & 45 & 3 & 19 & 6.85 & 0.34 & 0.68 \\
4B & 1.6 & 0.3 & 30 & 3 & 20 & 7.13 & 0.51 & 0.47 \\
 \\
\tableline
\end{tabular}
\end{center}
\tablenotetext{a}{see Figure 4 for model profiles.}
\end{table*}

\clearpage

\begin{table*}
\caption{Physical Quantities of Mass Flows in NGC 1333 IRAS 4}\label{tbl-4}

\begin{center}
\begin{tabular}{lcccc}
\tableline
Derived  & & & IRAS & IRAS \cr
Quantity & Abbreviation & (units) & 4A\tablenotemark{a} & 4B\tablenotemark{b} \cr
\tableline
\tableline
\\
{\it Infall}\tablenotemark{c}\/  : \cr
\tableline
 \\
Mass Accretion Rate & $\dot M_{in}$ & M\sun\ yr$^{-1}$ & 1.1 $\times$ 10$^{-4}$ & 3.7 $\times$ 10$^{-5}$\\ 
 \\
\tableline
 \\
{\it Outflow}\tablenotemark{d}\/  : \cr
\tableline
 \\
Mass & $M_{out}$ & M\sun\ & 6.9 $\times$ 10$^{-4}$ & 1.5 $\times$ 10$^{-4}$\\ 
Momentum & $P_{out}$ & M\sun\ km s$^{-1}$ & 1.6 $\times$ 10$^{-2}$ & 1.6 $\times$ 10$^{-3}$\\ 
Kinetic Energy & $E_{out}$ & erg & 4.9 $\times$ 10$^{42}$ & 3.2 $\times$ 10$^{41}$\\ 
Momentum Flux & $F_{H_{2}CO}$ & M\sun\ km s$^{-1}$ yr$^{-1}$ & 4.2 $\times$ 10$^{-5}$ & 3.2 $\times$ 10$^{-6}$\\ 
Mechanical Luminosity & $L_{mech}$ & L\sun\ & 1.3 $\times$ 10$^{-1}$ & 7.4 $\times$ 10$^{-3}$\\ 
 \\
\tableline
\end{tabular}
\end{center}
\tablenotetext{a}{southern blue lobe only.}
\tablenotetext{b}{includes contribution from eastern blue lobe, some of
which may originate from 4A\am; see text.}
\tablenotetext{c}{derived from very uncertain $r_{in}$; see text.}
\tablenotetext{d}{for dense gas in outflow only; assumes outflow inclination 
angle of 10\deg\ for both 4A and 4B.}
\end{table*}

\clearpage

\begin{table*}
\caption{N$_{2}$H$^{+}$ Line Properties Derived from Hyperfine Structure Fits}\label{tbl-5}

\begin{center}
\begin{tabular}{ccccccccccc}
\tableline

& & \multicolumn{4}{c}{4A}& \multicolumn{4}{c}{4B} \cr
Value & Unit & mean & rms & min & max & & mean & rms & min & max \cr
\tableline
\tableline
 \\
$T_{B}^{tot}$        & K                   & 7.8  & 2.1   & 3.6  & 13.  & & 12.  & 4.4    & 4.8  & 21.  \cr
$V_{LSR}$            & km s$^{-1}$         & 6.95 & 0.359 & 6.48 & 8.2  & & 6.98 & 0.0928 & 6.75 & 7.38 \cr
$\Delta V$           & km s$^{-1}$         & 0.49 & 0.17  & 0.21 & 0.83 & & 0.50 & 0.20   & 0.16 & 0.81 \cr
 \\
$N$(N$_{2}$H$^{+}$)  & 10$^{12}$ cm$^{-2}$ & 2.4  & 1.2   & 0.54 & 5.1  & & 3.9  & 1.9    & 0.77 & 7.7  \cr
\tableline
\end{tabular}
\end{center}
\end{table*}

\clearpage

\figcaption{Millimeter continuum and line interferometer maps of NGC 1333 IRAS
4.  Positions of continuum sources 4A, 4B, and 4B\am\ at $\lambda$ = 1.3 mm 
are denoted by {\it large black stars} while the position of continuum source 
4A\am\ from Looney et al.\/ (2000) is denoted by a {\it smaller black star}\/.  
The {\it black ellipses} placed in the bottom left of each panel represent the 
FWHMs of the respective synthesized beams.  Outer dashed contours represent 
the masking cutoffs at a specific gain level of the mosaics: 20\% for the IRAM 
PdBI, or 50\% for the NRO NMA data.  {\it a}\/) $\lambda$ = 1.3 mm continuum 
emission.  {\it Plain contours} start at 52.5 mJy beam$^{-1}$ (3 $\sigma$), 
increasing in 
steps of 2 $\sigma$ to 297.5 mJy beam$^{-1}$; the {\it dashed contour} denotes 
-52.5 mJy beam$^{-1}$.  {\it b}\/) $\lambda$ = 2.0 mm continuum emission.  
{\it Plain contours} start at 42.0 mJy beam$^{-1}$ (3 $\sigma$), increasing in 
steps of 2 $\sigma$ to 126.0 mJy beam$^{-1}$; the {\it dashed contour} denotes 
-42.0 mJy beam$^{-1}$.  {\it c}\/) $\lambda$ = 3.2 mm continuum emission. {\it 
Plain contours} start at 13.5 mJy beam$^{-1}$ (3 $\sigma$), increasing in 
steps of 2 $\sigma$ to 49.5 mJy beam$^{-1}$; dashed contour denotes -13.5 mJy 
beam$^{-1}$.  The {\it arrows} to the northeast and southwest of 4A+4A\am\ 
denote the directions of the large-scale CO outflow detected by Blake et al.\/ 
(1995) {\it d}\/) The integrated intensity of H$_{2}$CO 3$_{12}$--2$_{11}$
emission (``zeroth-moment"; see text).  The {\it bold contour} shows the 
half-maximum isophote (57 K km s$^{-1}$), while {\it plain contours} step up 
and down in increments of 1.5 $\sigma$ (10.5 K km s$^{-1}$) from this level.  
{\it e}\/) The integrated intensity of CS 3--2 emission.  The {\it bold 
contour} shows the half-maximum isophote (11.1 K km s$^{-1}$), while {\it plain 
contours} step up and down in increments of 1.5 $\sigma$ (4.5 K km s$^{-1}$) 
from this level.  {\it f}\/) The integrated intensity of N$_{2}$H$^{+}$ 1--0 
emission (over all 7 hyperfine components).  The {\it bold contour} shows the 
half-maximum isophote (7.6 K km s$^{-1}$), while {\it plain contours} step up 
and down in increments of 1.5 $\sigma$ (1.2 K km s$^{-1}$) from this level. 
\label{fig1}}

\figcaption{Grids of H$_{2}$CO 3$_{12}$-2$_{11}$ spectra at selected positions
at and around 4A and 4B.  In both grids, panels are spaced by 2\as\ in R.A. or 
Dec., $\sim$1 FWHM of the beam at 225.7 GHz, the H$_{2}$CO 3$_{12}$-2$_{11}$
rest frequency, so adjacent panels in the H$_{2}$CO grid are independent.  The 
dashed line represents the determined $V_{LSR}$ of the IRAS 4 system of 6.96
km s$^{-1}$.  {\it a) top grid}\/: Panels centered at the 4A position at 
$\lambda$ = 1.3 mm (R.A. (2000) = 03:29:10.5, Dec. (2000) = +31:13:31.4).  
{\it b) bottom grid}\/: Panels centered at the 4B position of 4B at $\lambda$ 
= 1.3 mm (R.A. (2000) = 03:29:12.0, Dec. (2000) = +31:13:08.2).  
\label{fig2}}

\figcaption{Grids of N$_{2}$H$^{+}$ 1--0 spectra at the same positions at 
and around 4A and 4B defined for Figure 2.  Due to the larger beam at 93.17
GHz, the N$_{2}$H$^{+}$ 1--0 rest frequency, the adjacent panels are not
independent, as in Figure 2.  In addition, the velocity range in each panel
here is wider relative that of to Figure 2, although panels in both Figures
are centered at the same velocity.
\label{fig3}}

\figcaption{H$_{2}$CO 3$_{12}$--2$_{11}$ spectra from the IRAM PdBI toward 
the positions of peak continuum intensity near 4A ({\it panel a}\/) and 4B 
({\it panel b}\/).  The {\it dotted line}\/ denotes the average systemic 
$V_{LSR}$ of $+$6.96 km s$^{-1}$.  In each panel, a {\it dashed line}\/ shows 
a spectrum resulting from a simple two-layer model of infall toward each 
object.  Respective model parameters are listed in Table 3.
\label{fig4}}

\figcaption{Channel maps of H$_{2}$CO 3$_{12}$--2$_{11}$ emission revealing
the underlying structure of the southern blue lobe of the 4A dense gas outflow. 
Each channel is the mean of 3 channels of the original data cube, and has a 
channel width of 0.3113 km s$^{-1}$.  The velocity of a given panel is shown 
at the upper left in km s$^{-1}$; only a fraction of channels blueward of the
systemic velocity are shown.  Continuum positions of 4A and 4A\am\ are denoted 
by {\it large stars}\/ and {\it small stars}\/ respectively.  {\it Solid 
contours}\/ begin at 3.5 K and increase in steps of 3.5 K, and {\it dashed 
contours}\/ begin at -3.5 K and decrease in steps of -3.5 K.  The {\it grey 
scale}\/ ranges from -3.5 K to 28.2 K.
\label{fig5}}

\figcaption{Channel maps of H$_{2}$CO 3$_{12}$--2$_{11}$ emission depicting 
the red and blue lobes of the 4B dense gas outflow.  Channel width, contours, 
grey scales, and symbols are the same as described for Figure 5, except {\it 
stars}\/ denote the continuum locations of 4B (near center) and 4B\am\ (to 
the left).  In addition, an equal number of channels redward and blueward of 
the systemic velocity of 4B are shown. 
\label{fig6}}

\figcaption{The spatial distribution of $T_{B}^{tot}$, $V_{LSR}$ and $\Delta
V$ of the N$_{2}$H$^{+}$ 1--0 line around IRAS 4A and 4B, obtained by 
fitting the 7 hyperfine components of the line simultaneously with the HFS 
fitting routine of CLASS.  Only positions where the peak intensity of
brightest hyperfine component was $\geq$5$\times$ the spectral baseline rms 
are shown, and data at other positions have been blanked.  The positions of 
continuum objects 4A, 4A\am, and 4B, are represented by {\it stars}, and are 
the same positions as denoted in Figure 1.
{\it a}\/) Distribution of $T_{B}^{tot}$ around 4A, where the color scale is 
defined from 0 K to 40 K, and contours begin at 2.5 K and continue in
increments of 2.5 K to 40.0 K.  
{\it b}\/) Distribution of $V_{LSR}$ around 4A, where the color scale 
is defined from 6.0 km s$^{-1}$ to 8.0 km s$^{-1}$, and contours begin at 6.2 
km s$^{-1}$ and increase in increments of 0.2 km s$^{-1}$ to 7.8 km s$^{-1}$.  
{\it c}\/) Distribution of $\Delta$V around 4A, where the color scale 
is defined from 0.1 km s$^{-1}$ to 1.5 km s$^{-1}$, and contours begin at 0.2
km s$^{-1}$ and increase in increments of 0.2 km s$^{-1}$ to 1.4 km s$^{-1}$.  
{\it d}\/) Distribution of $T_{B}^{tot}$ around 4B, with the same color scale 
and contour levels defined for panel a).
{\it e}\/) Distribution of $V_{LSR}$ around 4B, with the same color scale 
and contour levels defined for panel b).  
{\it f}\/) Distribution of $\Delta V$ around 4B, with the same color scale 
and contour levels defined for panel c). 
\label{fig7}}

\figcaption{Distribution of $\Delta V$ from N$_{2}$H$^{+}$ 1--0 emission
around 4A ({\it panel a}\/) and 4B ({\it panel b}\/) in 0.15 km s$^{-1}$ wide
bins as fractions of the total number of non-blanked pixels, i.e., 54 pixels 
for 4A and 12 pixels for 4B.  (The criterion for blanking is the same as for 
Figure 7.)  Values of $\Delta V$ are obtained by deconvolving the measured 
values by the 0.16 km s$^{-1}$ velocity resolution of the PdBI correlator.  
The {\it dashed line}\/ indicates the FWHM expected from the N$_{2}$H$^{+}$ 
1--0 line from thermal broadening only at $T_{K}$ = 40 K, the maximum $T_{K}$ 
estimated for the non-outflowing gas around 4A by B95.  The {\it dot-dashed 
line}\/ indicates the mean $\Delta V$ measured from spectra around each 
protostellar object.  \label{fig8}}

{
\noindent
Fig. A1.--- Spectral energy distributions of 4A, 4B, and 4B\am\ objects 
from interferometer continuum observations between $\lambda$ = 800 $\mu$m 
to $\lambda$ = 6 cm.  {\it Filled squares}, {\it filled triangles}, or {\it 
filled stars} show data obtained respectively from the IRAM PdBI, the NRO NMA, 
or the NRAO VLA for this paper.  The {\it open circles} show data at $\lambda$ 
= 800 $\mu$m from the JCMT-CSO Interferometer by Lay, Carlstrom, \& Hills 
(1995) for 4A and 4B.  {\it Open squares} show data at $\lambda$ = 1.3 cm 
and $\lambda$ = 3.6 cm from Mundy et al.\/ (1993) and {\it open diamonds} 
show data at $\lambda$ = 3.6 cm and $\lambda$ = 6.0 cm by Rodriguez et al.\/ 
(1999) for 4A and 4B all from the Very Large Array.  {\it Downward-pointing 
arrows} indicate 3 $\sigma$ upper limits.  {\it Error bars} are 1 $\sigma$ 
(see text and Table 2).  {\it Solid lines} indicate lines of slope $\alpha_{1}$ 
from resulting from linear least-squares fits to our flux densities of each 
object at 1.3 mm $<$ $\lambda$ $<$ 1.3 cm.  {\it Dashed lines} indicate lines 
of slope $\alpha_{2}$ resulting from similar fitting but using only our data 
at 1.3 mm $<$ $\lambda$ $<$ 3.2 mm.
} 

\clearpage

\begin{figure}
\vspace{7.25in}
\hspace{-0.25in}
\includegraphics{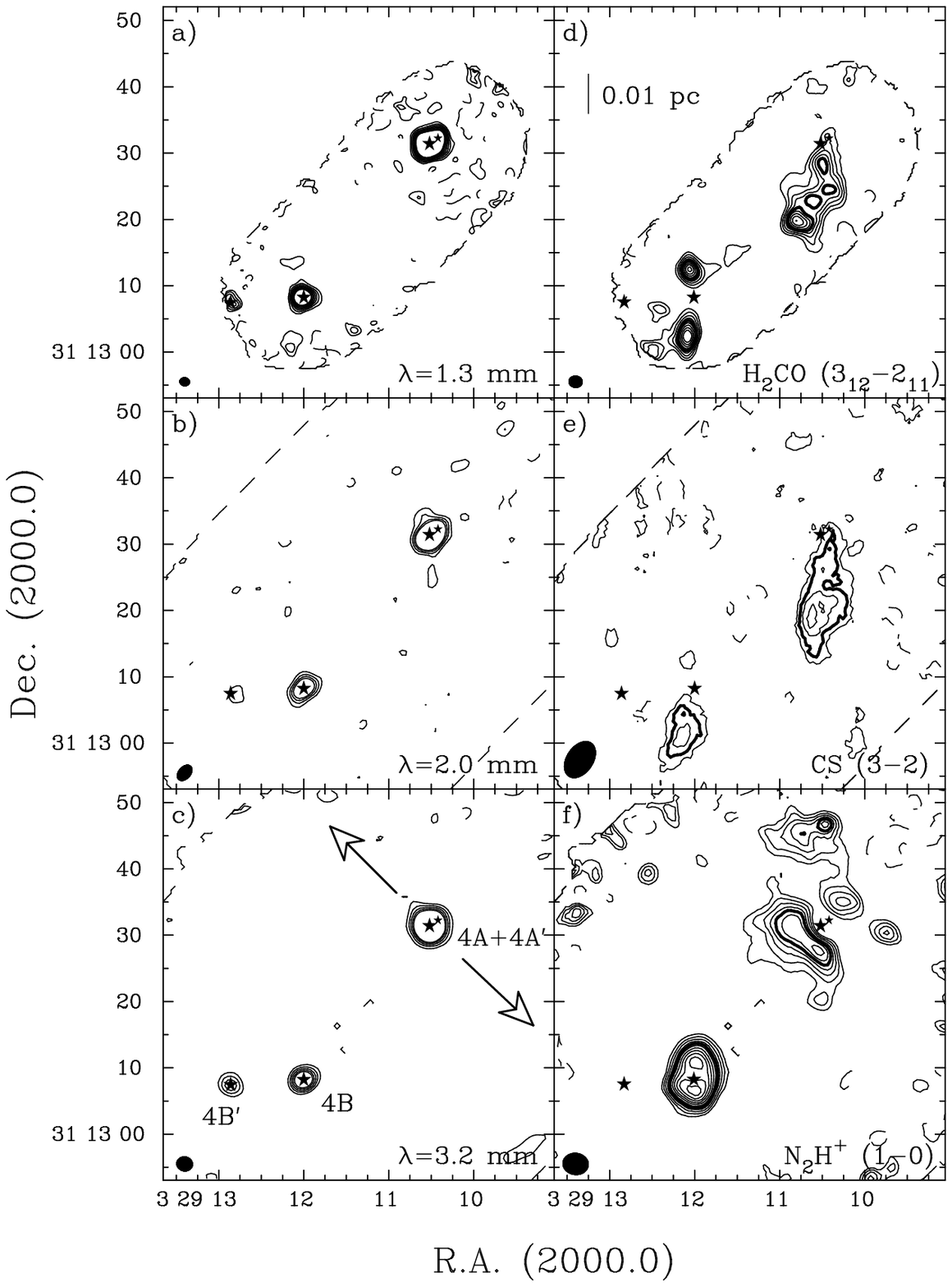}
\vspace{-2.25in}
\end{figure}

\begin{figure}
\vspace{7.25in}
\hspace{-0.25in}
\includegraphics{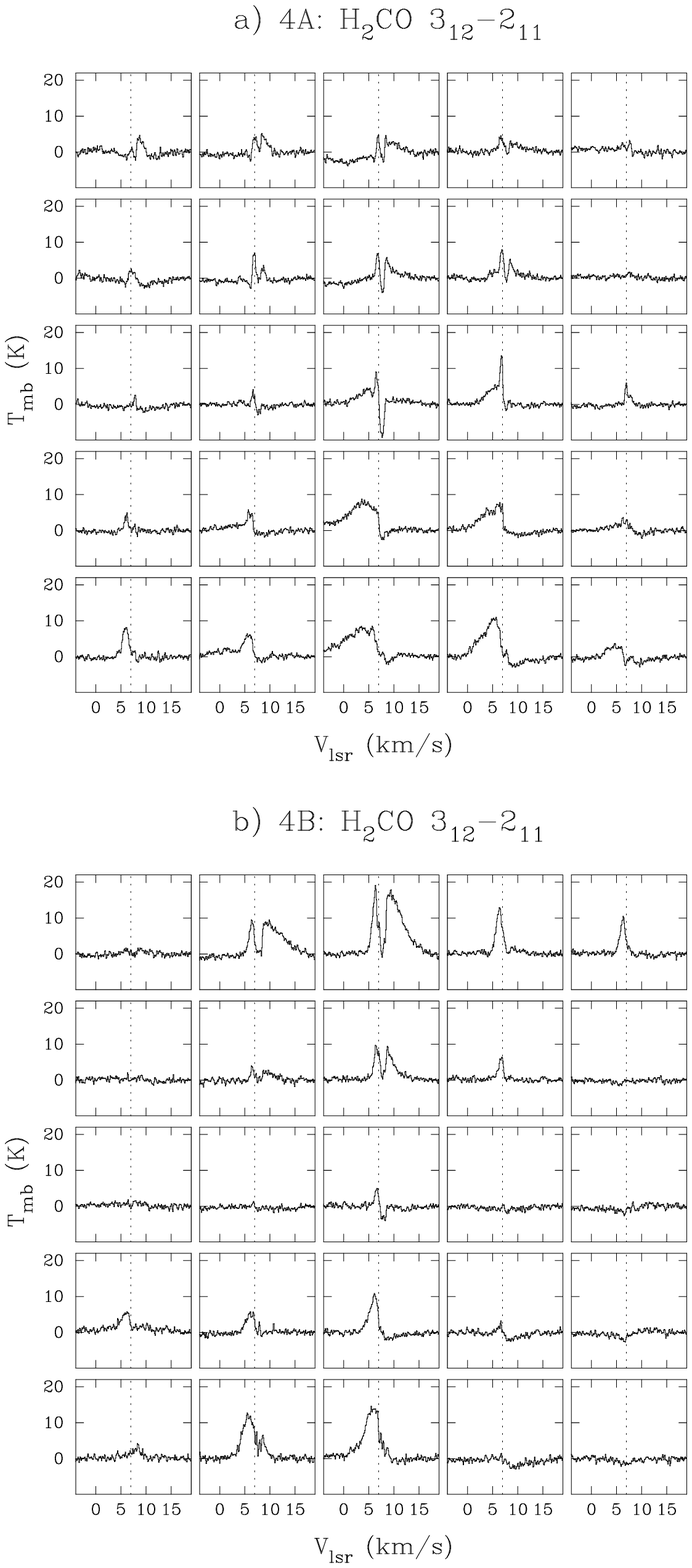}
\vspace{-2.25in}
\end{figure}

\begin{figure}
\vspace{7.25in}
\hspace{-0.25in}
\includegraphics{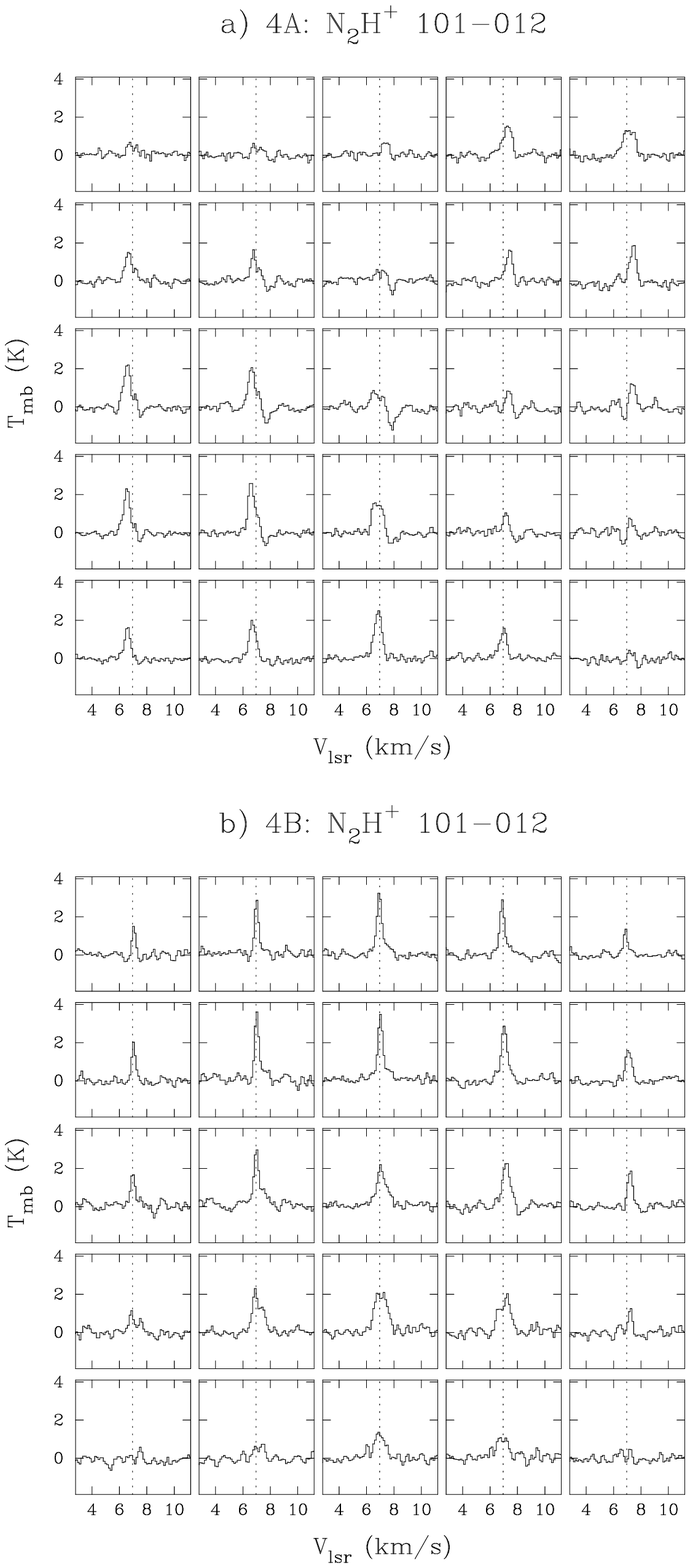}
\vspace{-2.25in}
\end{figure}

\begin{figure}
\vspace{7.25in}
\hspace{-0.25in}
\includegraphics{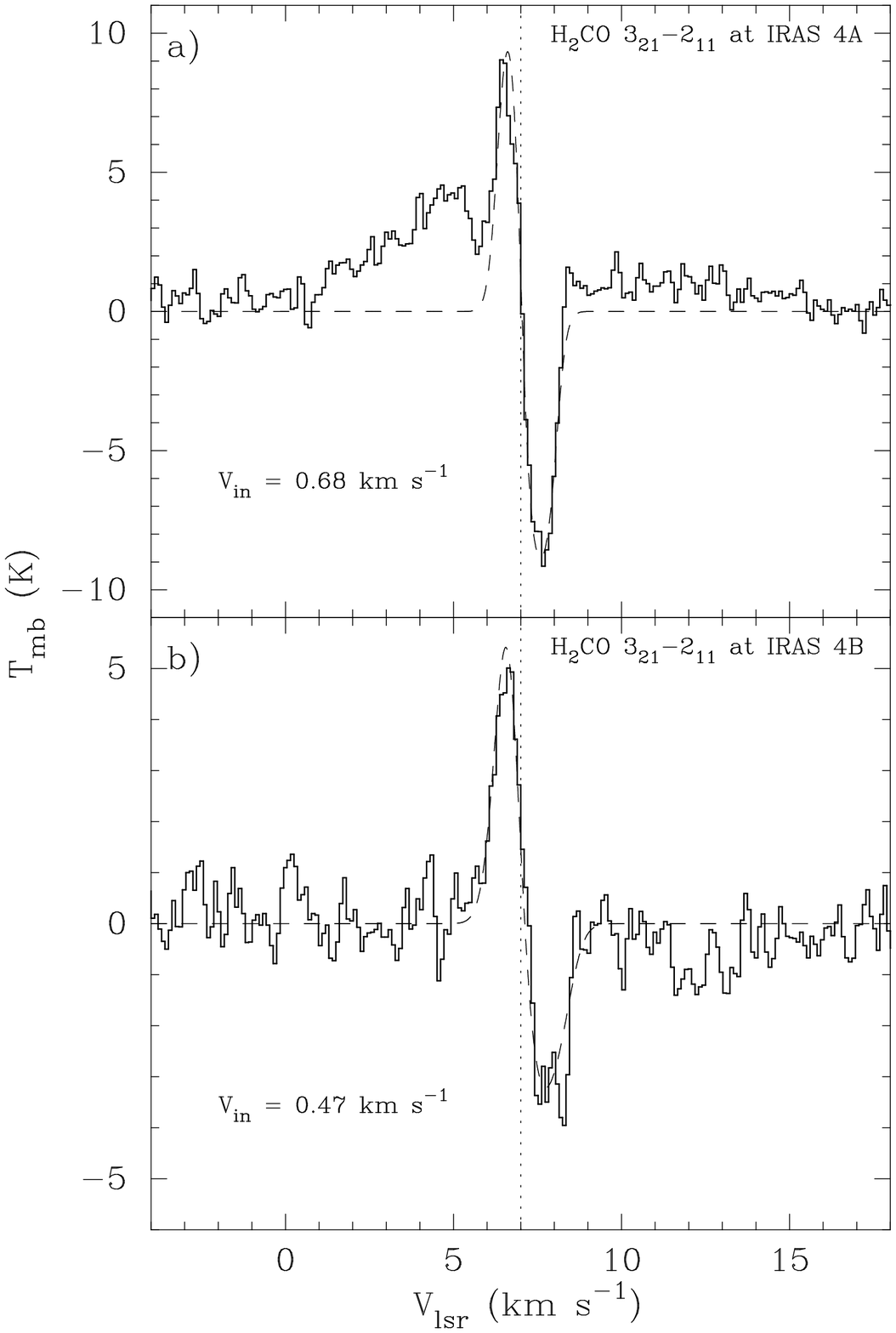}
\vspace{-2.25in}
\end{figure}

\begin{figure}
\vspace{7.25in}
\hspace{-0.25in}
\includegraphics{f5.eps}
\vspace{-2.25in}
\end{figure}

\begin{figure}
\vspace{7.25in}
\hspace{-0.25in}
\includegraphics{f6.eps}
\vspace{-2.25in}
\end{figure}

\begin{figure}
\vspace{7.25in}
\hspace{-0.25in}
\includegraphics{f7.eps}
\vspace{-2.25in}
\end{figure}

\begin{figure}
\vspace{7.25in}
\hspace{-0.25in}
\includegraphics{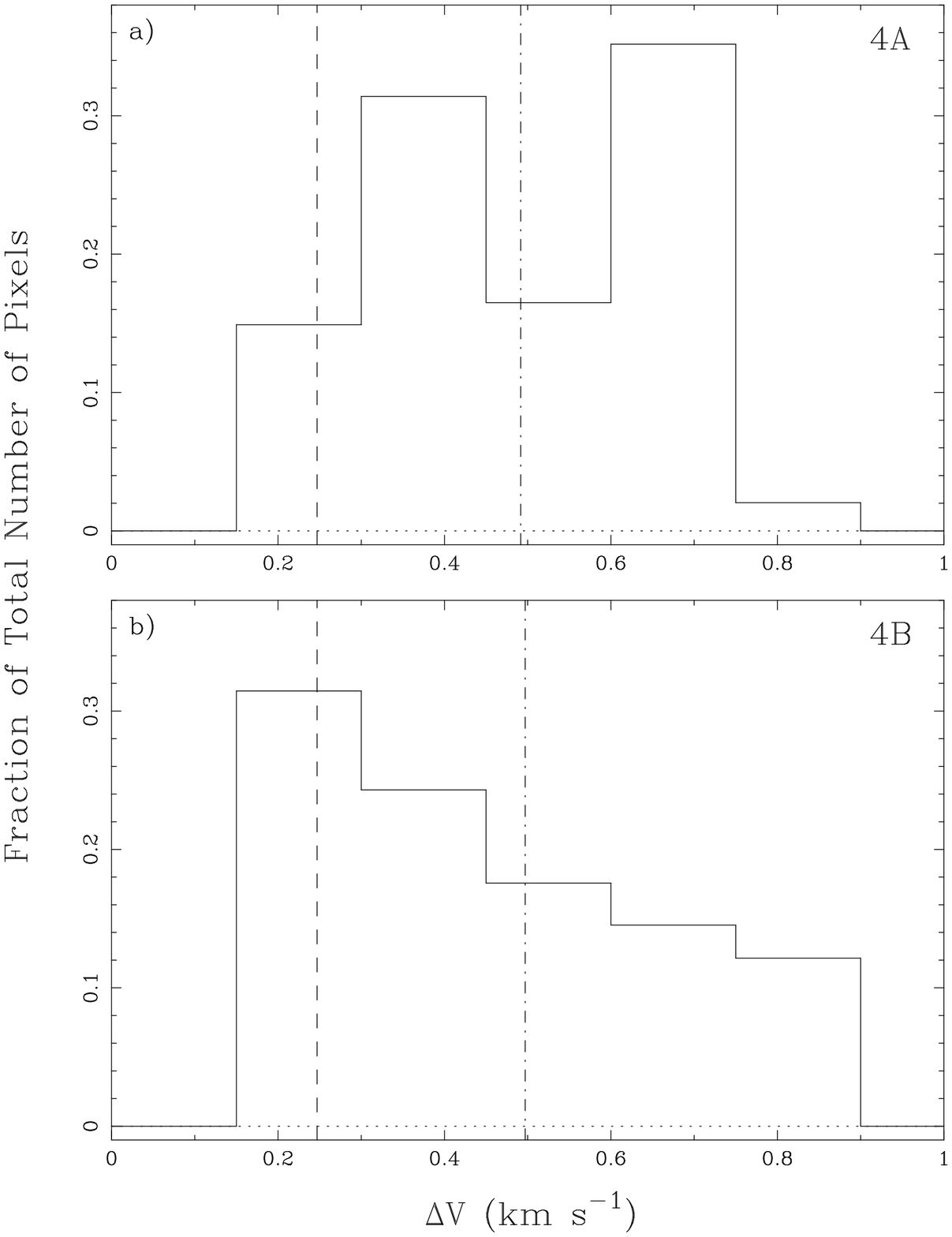}
\vspace{-2.25in}
\end{figure}

\begin{figure}
\vspace{7.25in}
\hspace{-0.25in}
\includegraphics{f9.eps}
\vspace{-2.25in}
\end{figure}

\end{document}